\documentclass{article}
\usepackage{iclr2027_conference,times}

\iclrfinalcopy

\usepackage[utf8]{inputenc}
\usepackage[T1]{fontenc}
\usepackage{hyperref}
\usepackage{url}
\usepackage{booktabs}
\usepackage{amsfonts}
\usepackage{microtype}
\usepackage{xcolor}

\usepackage{graphicx}
\usepackage{subcaption}
\usepackage{amsmath}
\usepackage{amssymb}
\usepackage{multirow}
\usepackage{pifont}
\usepackage{xspace}
\usepackage{bm}
\usepackage{wrapfig}

\newcommand{\cmark}{\ding{51}\xspace}
\newcommand{\xmark}{\ding{55}\xspace}

\title{Bayesian Speech Synthesizers Can Learn from Multiple Teachers}

\author{
  Ziyang Zhang\textsuperscript{1,2}\thanks{Corresponding author.} \quad
  Yifan Gao\textsuperscript{2} \quad
  Xuenan Xu\textsuperscript{2} \quad
  Baoxiang Li\textsuperscript{2} \quad
  Wen Wu\textsuperscript{2} \quad
  Chao Zhang\textsuperscript{1,2}\footnotemark[1] \\[4pt]
  {\normalfont \textsuperscript{1}Tsinghua University \quad
  \textsuperscript{2}Shanghai Artificial Intelligence Laboratory} \\[2pt]
  {\normalfont \texttt{ziyang-z24@mails.tsinghua.edu.cn}, \texttt{cz277@tsinghua.edu.cn}}
}

\begin{document}

\maketitle
\lhead{Preprint}

\begin{abstract}
  Text-to-speech (TTS) is a ``one-to-many'' problem: one text can be spoken in many valid ways, so its realizations form a distribution whose spread depends on the input. Most TTS models do not model this spread explicitly: they leave it implicit in the sampling process or fix it in advance. To model this distribution explicitly, we propose \textbf{BELLE} (\textbf{B}ayesian \textbf{e}vidential \textbf{l}earning with \textbf{l}anguag\textbf{e} modelling)\footnote{Audio samples and code are publicly available at \url{https://windowso.github.io/BELLE-demo/} and \url{https://github.com/OpenTSLab/BELLE}.}, which places a Bayesian prior over the mean and variance of each mel-spectrogram frame and predicts this prior from the input, while keeping the architecture and single-pass sampling of common continuous-valued autoregressive TTS models such as MELLE. Estimating this variance requires several recordings of each text, whereas standard TTS corpora provide one. We therefore construct two parallel-speech corpora: \textbf{ParallelVox} from real human recordings and \textbf{LS+6TTS}, which adds speech from six pre-trained TTS teachers to LibriSpeech (LS). Trained on these corpora, BELLE outperforms MELLE under the same training conditions, is competitive with larger TTS systems on the objective metrics of LibriSpeech test-clean, and supports streaming generation. Analyses of repeated human readings show that real speech has the properties the Bayesian prior assumes and that the variance predicted by BELLE follows the variation across these readings. Parallel speech improves BELLE, and synthetic parallel speech performs comparably to real parallel speech at the same volume and is far easier to scale.
\end{abstract}

\section{Introduction}
Text-to-Speech (TTS) synthesis is fundamentally a ``one-to-many'' mapping problem, because one text can be spoken in many valid ways. Different speakers read the same sentence with different timbre, prosody, and rhythm, and the same speaker also reads it differently from one repetition to the next. The valid realizations of one text therefore form a distribution, and how widely this distribution spreads depends on the text and its context. A TTS model should thus learn this distribution, including its spread for each input, rather than a single output.

Existing TTS models fall into three main families. Discrete codec-based autoregressive (AR) models \citep{borsos2023audiolm, wang2023neural} predict codec tokens, but tokenization introduces quantization error. Diffusion TTS models \citep{eskimez2024e2, chen2024f5, lee2025ditto} generate continuous features from noise through iterative denoising. Continuous-valued AR models predict mel-spectrogram frames one by one, either with diffusion or flow-matching heads \citep{liu2024autoregressive, wang2025felle, jia2025ditar} or, as in MELLE \citep{meng2024autoregressive}, with a Gaussian whose Kullback-Leibler (KL) loss is minimized when the variance equals one for every frame, regardless of the input. In these models, the spread of the ``one-to-many'' distribution is either left implicit in the sampling process or fixed to a constant, rather than modeled explicitly.

To model this distribution explicitly, we propose BELLE (\textbf{B}ayesian \textbf{e}vidential \textbf{l}earning with \textbf{l}anguag\textbf{e} modelling), a continuous-valued AR TTS model built on evidential deep learning \citep{amini2020deep}. Instead of fixing the variance, BELLE places a Bayesian prior over the mean and variance of each frame and predicts this prior from the input, so the spread of the distribution is estimated for each input rather than set in advance. This change only affects the layer that outputs each frame, so BELLE keeps the architecture and single-pass sampling of MELLE.

Implementing this Bayesian framework, however, introduces a statistical challenge: accurate variance estimation theoretically requires multiple observations per input, whereas standard TTS datasets typically provide only a single recording per text. Two complementary sources of ``one-to-many'' supervision are therefore explored. On the real side, we construct \textbf{ParallelVox}, a human parallel-speech corpus assembled from five public datasets, in which every text prompt is paired with multiple recordings. On the synthetic side, LibriSpeech (LS; \citealp{panayotov2015librispeech}) is augmented with parallel speech from six pre-trained TTS teachers (denoted \textbf{LS+6TTS}), which gives each text multiple realizations.

Trained on these parallel-speech corpora, BELLE outperforms MELLE under the same training conditions, is competitive with larger TTS systems on the objective metrics of LibriSpeech test-clean, and supports streaming generation. Analyses of repeated human readings of the same text show that real speech has the properties this Bayesian prior assumes, and that the variance predicted by BELLE changes with the input and follows the variation across these readings. On the data side, parallel speech improves BELLE over single-recording data of similar volume, and synthetic parallel speech performs comparably to real parallel speech at the same volume and is far easier to scale.

\textbf{Summary of Contributions.}
\begin{itemize}
  \item We propose BELLE, a TTS model that explicitly models the ``one-to-many'' distribution with evidential deep learning, predicting a Bayesian prior over the mean and variance of each frame.
  \item We construct two parallel-speech corpora with multiple realizations per text: ParallelVox from real human recordings and LS+6TTS from six pre-trained TTS teachers. At the same volume, synthetic parallel speech performs comparably to real parallel speech and is far easier to scale.
  \item Trained on these corpora, BELLE outperforms MELLE under the same training conditions, is competitive with larger TTS systems on the objective metrics of LibriSpeech test-clean, and supports streaming generation.
\end{itemize}

\section{Related Work}

\textbf{Continuous-valued TTS.} Continuous-valued TTS models generate acoustic features such as mel-spectrograms or variational autoencoder (VAE) latents without a discrete tokenizer.
Non-autoregressive (NAR) models such as E2 TTS \citep{eskimez2024e2} and F5-TTS \citep{chen2024f5} generate the whole utterance through iterative flow-matching refinement.
AR models generate one frame at a time and differ in how each frame is sampled: MELLE \citep{meng2024autoregressive} samples from a Gaussian trained with a KL loss toward a unit-variance target, several models attach diffusion or flow-matching heads \citep{liu2024autoregressive, wang2025felle, jia2025ditar}, and others predict a Gaussian or Gaussian-mixture distribution over VAE latents \citep{lin2025continuous, zhu2024autoregressive}.
These VAE-based models learn the distribution of each frame from one recording per text, with a likelihood loss or a KL loss toward the VAE posterior of that recording.
BELLE keeps single-pass sampling of each frame, but places a Bayesian prior over the mean and variance of the frame, and when parallel speech is available, trains this prior on all realizations of a text.

\textbf{Streaming TTS.} Streaming TTS systems output speech while the input text is still arriving.
Several systems pair a language model over discrete speech tokens with a flow-matching decoder \citep{du2024cosyvoice, sheng2025syncspeech} or interleave text and speech tokens \citep{yang2024interleaved}, while SMLLE \citep{sun2025zero} and StreamMel \citep{wang2025streammel} generate continuous features directly.
BELLE-stream also generates mel-spectrograms within one continuous-valued AR model by interleaving text and speech chunks, and it uses the same evidential sampling as BELLE, so the Bayesian prior is predicted for every frame during streaming.

\textbf{Evidential Learning and Bayesian Neural Networks.} Bayesian deep learning places priors either on network weights, as in Bayesian Neural Networks (BNNs) \citep{mackay1992practical, neal1996bayesian}, or on the parameters of the likelihood, as in Evidential Deep Learning (EDL) \citep{sensoy2018evidential, amini2020deep}.
BNNs approximate the weight posterior with variational inference \citep{graves2011practical, blundell2015weight} or Monte Carlo dropout \citep{gal2016dropout}, so a prediction typically requires several forward passes, and in AR TTS this extra cost is paid at every frame.
EDL instead predicts the parameters of a conjugate prior, such as the Normal-Inverse-Gamma (NIG) prior over the mean and variance of a Gaussian, and marginalizes over them in closed form within one forward pass, which places it in the Empirical Bayes framework (Appendix~\ref{app:bayesian_clarification}).
BELLE applies EDL to every frame of a continuous-valued AR TTS model, so the prior over each frame is predicted at the cost of a single forward pass.

\section{Mel-based Autoregressive TTS}
\subsection{Problem Formulation}

Mel-based AR TTS models formulate synthesis as next-frame prediction. Given an input text sequence \( \boldsymbol{x} = [x_1, \ldots, x_L] \), the model generates a mel-spectrogram \( \boldsymbol{y} = [\boldsymbol{y}_1, \ldots, \boldsymbol{y}_T] \), where each frame \( \boldsymbol{y}_t \in \mathbb{R}^{D} \) is the spectrum at time step \(t\) and \(D\) is the number of mel-frequency bands. Each frame \(\boldsymbol{y}_t\) is generated conditioned on the text \(\boldsymbol{x}\) and the previous frames \(\boldsymbol{y}_{<t} = [\boldsymbol{y}_1, \ldots, \boldsymbol{y}_{t-1}]\), so the distribution of \(\boldsymbol{y}\) factorizes as
\begin{equation}
\label{eq:ar_factorization}
  p(\boldsymbol{y} \mid \boldsymbol{x};\theta) = \prod\nolimits_{t=1}^{T} p(\boldsymbol{y}_t \mid \boldsymbol{y}_{<t}, \boldsymbol{x};\theta),
\end{equation}
where \(\theta\) denotes the parameters of the neural network that models each conditional distribution.

In \textbf{streaming generation}, speech is output while the input text is still arriving. Both the input text and the target mel-spectrogram are therefore split into \(K\) consecutive chunks, \( \boldsymbol{X} = [\boldsymbol{x}^{(1)}, \ldots, \boldsymbol{x}^{(K)}] \) and \( \boldsymbol{Y} = [\boldsymbol{y}^{(1)}, \ldots, \boldsymbol{y}^{(K)}] \). Unlike Eq.~\eqref{eq:ar_factorization}, where every frame is conditioned on the full text, the \(k\)-th audio chunk \(\boldsymbol{y}^{(k)}\) is conditioned only on the previous audio chunks \(\boldsymbol{y}^{(<k)}\) and on the text chunks received so far, \(\boldsymbol{x}^{(\le k)}\):
\begin{equation}
\label{eq:stream_factorization}
  p(\boldsymbol{Y} \mid \boldsymbol{X}; \theta)
  = \prod\nolimits_{k=1}^{K} p\big(\boldsymbol{y}^{(k)} \mid \boldsymbol{y}^{(<k)}, \boldsymbol{x}^{(\le k)}; \theta\big).
\end{equation}

\subsection{Model Architecture}
\label{sec:architecture}

\begin{figure*}
  \centering
  \includegraphics[width=1\linewidth]{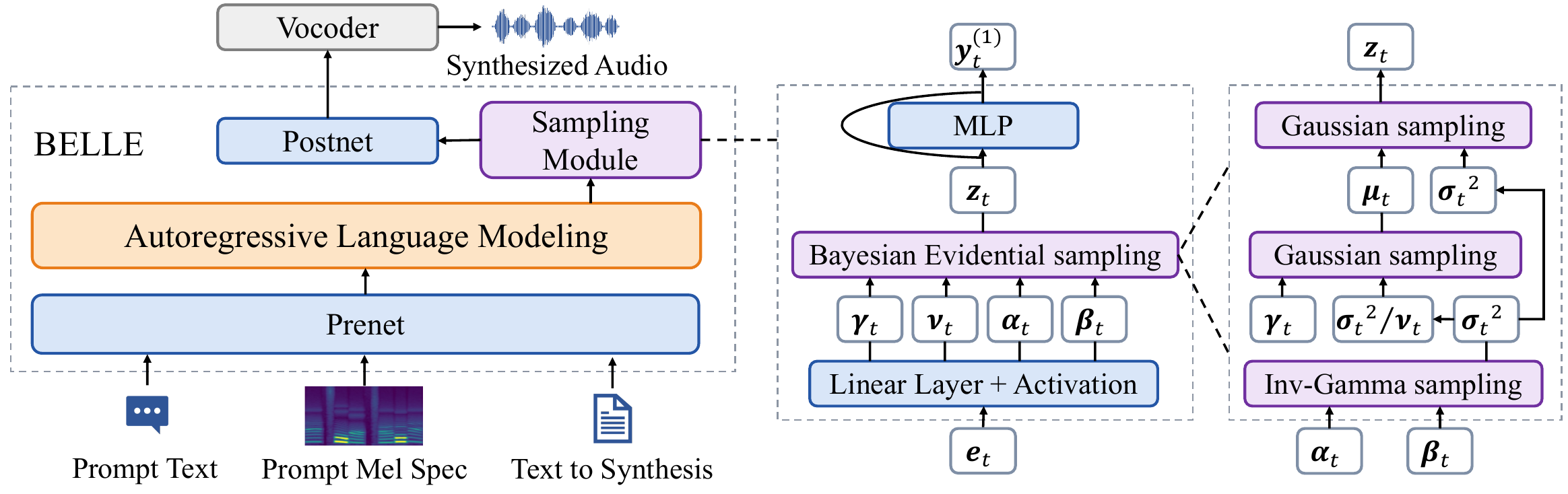}
  \caption{Overview of BELLE. Left: the mel-based AR TTS architecture (Sec.~\ref{sec:architecture}). Right: the Sampling Module, which predicts a Bayesian prior over the mean and variance of each frame and samples the frame from it (Sec.~\ref{sampling}).}
  \label{fig:BELLE}
\end{figure*}
As shown in Fig.~\ref{fig:BELLE}, a general mel-based AR TTS model consists of a \textbf{Prenet}, an \textbf{AR LM} (AR language model), a \textbf{Sampling Module}, a \textbf{Stop-Prediction Module}, a \textbf{Postnet}, and a \textbf{Vocoder}. The Prenet projects the text tokens \(\boldsymbol{x}\) and the previous mel frames to the hidden dimension of the AR LM, a decoder-only Transformer. At each step \(t\), the AR LM attends to the text and to the frames before \(t\) and outputs a hidden representation \(\boldsymbol{e}_t\), from which frame \(t\) is generated. The Sampling Module projects \(\boldsymbol{e}_t\) to the parameters of a distribution over the frame, which is an NIG distribution in BELLE (Sec.~\ref{sampling}), and draws a sample \(\boldsymbol{z}_t\) from it. Because \(\boldsymbol{z}_t\) is a random sample, it carries sampling noise, so a denoising network, implemented as a multilayer perceptron (MLP), refines \(\boldsymbol{z}_t\) into the frame \(\boldsymbol{y}_t^{(1)}\) (Fig.~\ref{fig:BELLE}, right). Because the model generates continuous mel frames rather than discrete tokens, it cannot end generation by emitting an end-of-sequence token as discrete AR models do. Instead, as in TransformerTTS and SpeechT5 \citep{li2019transformertts, ao2021speecht5}, the Stop-Prediction Module, a linear layer with sigmoid activation applied to \(\boldsymbol{e}_t\), outputs a stop score in \([0,1]\), and generation halts once the score exceeds a threshold. Following Tacotron2 and TransformerTTS (\citealp{shen2018tacotron2}; \mbox{\citealp{li2019transformertts}}), the Postnet then adds a convolutional residual to the generated mel-spectrogram \(\boldsymbol{y}^{(1)}\), yielding the refined mel-spectrogram \(\boldsymbol{y}^{(2)}\), which the Vocoder converts into a waveform.

\begin{wrapfigure}{r}{0.5\textwidth}
    \centering
    \vspace{-12pt}
    \includegraphics[width=\linewidth]{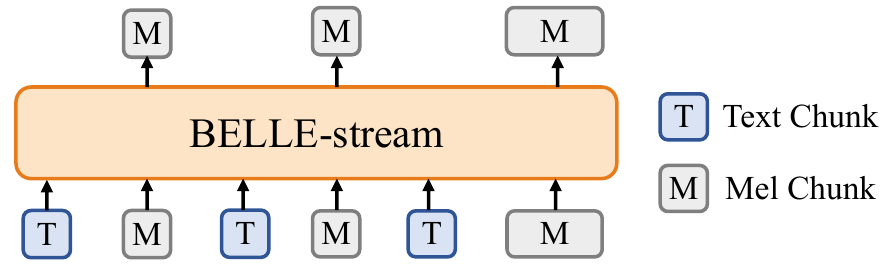}
    \caption{Interleaved text and mel-spectrogram chunks in BELLE-stream.}
    \label{fig:belle_stream}
    \vspace{-10pt}
\end{wrapfigure}
BELLE-stream uses the same architecture and implements Eq.~\eqref{eq:stream_factorization} by interleaving text and audio chunks as the input of the AR LM under a causal mask (Fig.~\ref{fig:belle_stream}). Each text chunk \(\boldsymbol{x}^{(k)}\) contains \(S_{\mathrm{text}}\) tokens and each audio chunk \(\boldsymbol{y}^{(k)}\) contains \(S_{\mathrm{audio}}\) frames. \(S_{\mathrm{audio}}\) is kept small so that each audio chunk is generated after enough text has been received. As a result, the generated speech lags behind the text, and when the last text chunk arrives, more than \(S_{\mathrm{audio}}\) frames typically remain to be generated. The final audio chunk therefore has no fixed length: generation continues until the Stop-Prediction Module halts it, so this chunk is longer than the preceding ones.

\section{Bayesian Evidential Learning}
\label{sampling}

\subsection{Evidential Modeling of Mel Frames}
\label{EDL}

EDL places a prior over the parameters of the likelihood and lets the network predict the parameters of this prior.
In BELLE, each mel-spectrogram frame \(\boldsymbol{y}_t \in \mathbb{R}^D\) is modeled as a Gaussian with unknown mean \(\boldsymbol{\mu}_t\) and variance \(\boldsymbol{\sigma}_t^{2}\), and an NIG prior is placed over them:
\begin{equation}
  \boldsymbol{y}_t \sim \mathcal{N}(\boldsymbol{\mu}_t, \boldsymbol{\sigma}_t^{2}),\quad
  \boldsymbol{\mu}_t \sim \mathcal{N}\left(\boldsymbol{\gamma}_t,\frac{\boldsymbol{\sigma}_t^{2}}{\boldsymbol{\nu}_t}\right),\quad
  \boldsymbol{\sigma}_t^{2} \sim \Gamma^{-1}(\boldsymbol{\alpha}_t, \boldsymbol{\beta}_t),
  \label{eq:nig_prior_mel_vector}
\end{equation}
where the NIG parameters \(\boldsymbol{\gamma}_t, \boldsymbol{\nu}_t, \boldsymbol{\alpha}_t, \boldsymbol{\beta}_t \in \mathbb{R}^D\) satisfy \(\nu_{t,d} > 0\), \(\alpha_{t,d} > 1\), and \(\beta_{t,d} > 0\) for every dimension \(d \in \{1,\ldots,D\}\).
All distributions act element-wise, so the \(D\) mel bands are modeled independently (Appendix~\ref{app:dependency_modeling} discusses this choice).

Integrating out \(\boldsymbol{\mu}_t\) and \(\boldsymbol{\sigma}_t^{2}\) gives the predictive distribution of the frame, which is a Student-\(t\) distribution in each dimension:
\begin{equation}
  p(\boldsymbol{y}_t \mid \boldsymbol{\gamma}_t,\boldsymbol{\nu}_t,\boldsymbol{\alpha}_t,\boldsymbol{\beta}_t)
  = \mathrm{St}\!\left(\boldsymbol{y}_t;\,\boldsymbol{\gamma}_t,\,\frac{\boldsymbol{\beta}_t \odot (1+\boldsymbol{\nu}_t)}{\boldsymbol{\nu}_t \odot \boldsymbol{\alpha}_t},\, 2\boldsymbol{\alpha}_t\right),
  \label{eq:student_t_predictive_mel_vector}
\end{equation}
where \(\odot\) denotes element-wise multiplication.

\subsection{From Gaussian to Evidential Sampling}
\label{sec:evidential_sampling}

In MELLE \citep{meng2024autoregressive}, the Sampling Module predicts a diagonal Gaussian from the hidden representation \(\boldsymbol{e}_t\) with a linear projection,
\begin{equation}
\label{eq:melle_param}
    \left[\boldsymbol{\mu}_t, \log \boldsymbol{\sigma}_t^2\right] = \boldsymbol{W}_{\text{G}}\boldsymbol{e}_t + \boldsymbol{b}_{\text{G}},
\end{equation}
and draws \(\boldsymbol{z}_t = \boldsymbol{\mu}_t + \boldsymbol{\sigma}_t \odot \boldsymbol{\epsilon}\) with \(\boldsymbol{\epsilon} \sim \mathcal{N}(\boldsymbol{0}, \boldsymbol{I})\), a reparameterization that keeps the sampling step differentiable.
This Gaussian is trained with a KL divergence toward a target distribution centered at the ground-truth frame \(\boldsymbol{y}_t\) with unit variance:
\begin{equation}
\label{eq:melle_kl}
\begin{aligned}
    \mathcal{L}_{\text{KL}} &= \sum\nolimits_{t=1}^{T} D_{\text{KL}}\left( \mathcal{N}(\boldsymbol{\mu}_t, \boldsymbol{\sigma}_t^2 \boldsymbol{I}) \parallel \mathcal{N}(\boldsymbol{y}_t, \boldsymbol{I}) \right) \\
    &= \frac{1}{2} \sum\nolimits_{t=1}^{T} \left( \|\boldsymbol{\mu}_t - \boldsymbol{y}_t\|^2_2 + \sum\nolimits_{d=1}^{D} \left(\sigma_{t,d}^2 - \log \sigma_{t,d}^2\right) - D \right).
\end{aligned}
\end{equation}
The first term is a regression loss on the mean. Each variance term \(\sigma_{t,d}^2 - \log \sigma_{t,d}^2\) is minimized at \(\sigma_{t,d}^2 = 1\) independently of \(\boldsymbol{e}_t\), so the KL loss pulls the variance of every frame toward the same constant regardless of the input.

BELLE replaces this Gaussian with the NIG prior of Sec.~\ref{EDL}, as shown in the right part of Fig.~\ref{fig:BELLE}. A linear layer predicts the NIG parameters from \(\boldsymbol{e}_t\):
\begin{equation}
  \left[\boldsymbol{\gamma}_t, \boldsymbol{\nu}_t, \boldsymbol{\alpha}_t, \boldsymbol{\beta}_t\right]
  = \text{Activation}(\boldsymbol{W}_{\text{NIG}}\boldsymbol{e}_t + \boldsymbol{b}_{\text{NIG}}),
\end{equation}
where \(\text{Activation}(\cdot)\) applies softplus to \(\boldsymbol{\nu}_t\) and \(\boldsymbol{\beta}_t\) and a shifted softplus to \(\boldsymbol{\alpha}_t\), so that the constraints in Sec.~\ref{EDL} hold.
Each frame is then sampled hierarchically, following the generative process of Eq.~\eqref{eq:nig_prior_mel_vector}:
\begin{equation}
  \boldsymbol{\sigma}_{t}^{2} \sim \Gamma^{-1}(\boldsymbol{\alpha}_{t}, \boldsymbol{\beta}_{t}),\quad
  \boldsymbol{\mu}_{t} \sim \mathcal{N}\left(\boldsymbol{\gamma}_{t}, \frac{\boldsymbol{\sigma}_{t}^{2}}{\boldsymbol{\nu}_{t}}\right),\quad
  \boldsymbol{z}_{t} \sim \mathcal{N}\left(\boldsymbol{\mu}_{t}, \boldsymbol{\sigma}_{t}^{2}\right).
  \label{eq: invgamma sample}
\end{equation}
The resulting \(\boldsymbol{z}_t\) follows the Student-\(t\) distribution of Eq.~\eqref{eq:student_t_predictive_mel_vector}, whose scale is determined by \(\boldsymbol{\nu}_t\), \(\boldsymbol{\alpha}_t\), and \(\boldsymbol{\beta}_t\). Because these parameters are predicted from \(\boldsymbol{e}_t\), the spread of \(\boldsymbol{z}_t\) changes with the input.
In place of the KL loss of Eq.~\eqref{eq:melle_kl}, the NIG parameters are trained with the evidential loss, summed over frames and mel bands:
\begin{equation}
\label{eq:edl_loss_mel_vector}
\begin{aligned}
  \mathcal{L}_{\text{edl}} &= \sum\nolimits_{t=1}^{T} \sum\nolimits_{d=1}^{D} \left( \mathcal{L}_{\text{NLL}} + \lambda\,\mathcal{L}_{\text{R}} \right), \\
  \mathcal{L}_{\text{NLL}} &= \frac{1}{2}\log\frac{\pi}{\nu} - \alpha\log\Omega + \left(\alpha+\frac{1}{2}\right)\log\left(\nu(y-\gamma)^2+\Omega\right) + \log\frac{\Gamma(\alpha)}{\Gamma(\alpha+\frac{1}{2})}, \\
  \mathcal{L}_{\text{R}} &= |y-\gamma|\,(2\nu+\alpha),
\end{aligned}
\end{equation}
where the indices \(t,d\) are omitted inside the sum, \(\Omega = 2\beta(1+\nu)\), and \(\Gamma(\cdot)\) is the gamma function. \(\mathcal{L}_{\text{NLL}}\) is the negative log-likelihood (NLL) of the ground-truth frame under the Student-\(t\) distribution of Eq.~\eqref{eq:student_t_predictive_mel_vector}. The regularizer \(\mathcal{L}_{\text{R}}\) penalizes the evidence \(2\nu + \alpha\), which measures the confidence of the prior, in proportion to the error \(|y - \gamma|\), so that the prior becomes less confident on frames that are predicted poorly. Unlike Eq.~\eqref{eq:melle_kl}, this loss contains no term that pulls the variance toward a fixed value, so the spread of the predictive distribution is fit to the observed frames. Appendix~\ref{app:edl} gives the derivations.

\subsection{Training Objectives}
\label{subsec:training_objectives_belle}

BELLE is trained end-to-end with four losses:
\begin{equation}
  \mathcal{L} = \mathcal{L}_{\text{reg}} + \lambda_{\text{samp}}\mathcal{L}_{\text{samp}} + \lambda_{\text{flux}}\mathcal{L}_{\text{flux}} + \mathcal{L}_{\text{stop}},
  \label{eq:total_loss_belle}
\end{equation}
where \(\lambda_{\text{samp}}\) and \(\lambda_{\text{flux}}\) weight the sampling and flux losses (values in Appendix~\ref{app: trainingdetails}).

The regression loss \(\mathcal{L}_{\text{reg}}\) applies L1 and L2 losses between the ground-truth frames \(\boldsymbol{y}^{\text{gt}}\) and both the output of the denoising MLP, \(\boldsymbol{y}^{(1)}\), and the output of the Postnet, \(\boldsymbol{y}^{(2)}\):
\begin{equation}
  \mathcal{L}_{\text{reg}} = \sum\nolimits_{j=1}^{2}\left(\lVert\boldsymbol{y}^{\text{gt}} - \boldsymbol{y}^{(j)}\rVert_{1} + \lVert\boldsymbol{y}^{\text{gt}} - \boldsymbol{y}^{(j)}\rVert_{2}^{2}\right).
  \label{eq:regression_loss_belle_rhs}
\end{equation}
The sampling loss \(\mathcal{L}_{\text{samp}}\) trains the Sampling Module. It is the evidential loss \(\mathcal{L}_{\text{edl}}\) of Eq.~\eqref{eq:edl_loss_mel_vector} in BELLE and the KL loss of Eq.~\eqref{eq:melle_kl} in MELLE.

The spectrogram flux loss \(\mathcal{L}_{\text{flux}}\) rewards a large L1 distance between the predicted location \(\boldsymbol{\gamma}_t\) of frame \(t\) and the previous ground-truth frame \(\boldsymbol{y}_{t-1}^{\text{gt}}\), which penalizes predictions that copy the previous frame:
\begin{equation}
  \mathcal{L}_{\text{flux}} = -\sum\nolimits_{t=2}^{T}\lVert\boldsymbol{\gamma}_t - \boldsymbol{y}_{t-1}^{\text{gt}}\rVert_1.
  \label{eq:flux_loss_belle}
\end{equation}
MELLE applies the same loss to its Gaussian mean \(\boldsymbol{\mu}_t\) in place of \(\boldsymbol{\gamma}_t\).

The stop loss \(\mathcal{L}_{\text{stop}}\) is a binary cross-entropy loss on the output of the Stop-Prediction Module (Sec.~\ref{sec:architecture}). Only the last frame of each utterance is a positive sample, so its term is weighted by 500.

When a text \(\boldsymbol{x}\) has \(N \geq 1\) realizations \(\boldsymbol{y}^{\text{gt}}_1, \dots, \boldsymbol{y}^{\text{gt}}_N\), where \(N\) may differ across texts, each realization is used as a separate ground truth\label{sec: multi teacher}. With \(\mathcal{L}(\boldsymbol{y}^{\text{gt}}_i)\) denoting the loss of Eq.~\eqref{eq:total_loss_belle} computed against \(\boldsymbol{y}^{\text{gt}}_i\), the training loss for \(\boldsymbol{x}\) is the weighted sum
\begin{equation}
  \mathcal{L}_{\boldsymbol{x}} = \sum\nolimits_{i=1}^{N} w_i \mathcal{L}(\boldsymbol{y}^{\text{gt}}_i),\quad \text{with}\quad \sum\nolimits_{i=1}^{N} w_i = 1,\quad w_i \geq 0.
  \label{eq:one_to_many}
\end{equation}
Using all \(N\) realizations of \(\boldsymbol{x}\) as targets gives more direct evidence for the spread of Sec.~\ref{sec:evidential_sampling} than a single recording provides. Appendix~\ref{app:ar_realizations} shows that this objective is the likelihood of the \(N\) realizations under the AR factorization, although each realization is teacher-forced on its own previous frames.

\section{Parallel Speech Corpora}
\label{sec:data}

To help BELLE estimate the spread of the ``one-to-many'' distribution, two parallel-speech corpora with multiple realizations per text are constructed: ParallelVox from real human recordings and LS+6TTS from TTS synthesis.

\subsection{ParallelVox: Real Parallel Speech}
\label{sec:multivox}

\textbf{ParallelVox} merges five public English corpora in which the same text is recorded several times, either by different speakers or by one speaker on different occasions: LibriHeavy \citep{kang2024libriheavy}, the Parallel Audiobook Corpus, VCTK \citep{yamagishi2019vctk}, CMU ARCTIC \citep{kominek2004cmu} with L2-Arctic \citep{zhao2018l2arctic}, and the parallel subset of Mozilla Common Voice \citep{ardila2020common}. Each recording is transcribed with Whisper-large-v2 \citep{radford2023robust} and dropped if its word error rate (WER) exceeds 10\%, and a text is kept only if at least two of its recordings remain. Texts whose recordings all come from one speaker form the single-speaker repeat test set (22~h, 250 speakers, 2.1 recordings per text on average), and the remaining texts form the multi-speaker training set (881~h, 3{,}406 speakers, 2.8 recordings per text on average). The test set records how one speaker varies across repetitions of the same text, so it serves as the human reference for the diversity evaluation (Sec.~\ref{sec: evaluation}). On the training set, all recordings of a text are weighted equally in Eq.~\eqref{eq:one_to_many}. Appendix~\ref{app:multivox} gives the source statistics and the filtering details.

\subsection{LS+6TTS: Synthetic Parallel Speech}
\label{sec: training data}

Real parallel corpora are costly to extend: ParallelVox reaches 881~h only after five corpora are merged. \textbf{LS+6TTS} instead adds synthetic realizations to a corpus with one recording per text. The LibriSpeech \citep{panayotov2015librispeech} training set is trimmed with voice activity detection and restricted to utterances of 0.5 to 14~s, which gives 706~h (denoted \textbf{LS}). Six pre-trained TTS teachers, CosyVoice2 \citep{du2024cosyvoice}, IndexTTS \citep{deng2025indextts}, SparkTTS \citep{wang2025spark}, F5TTS \citep{chen2024f5}, MaskGCT \citep{wang2024maskgct}, and XTTS-v2 \citep{casanova2024xtts}, each synthesize every text, so each text has \(N=7\) realizations and the corpus totals 4{,}817~h. Appendix~\ref{app: training data} gives the processing details.

In Eq.~\eqref{eq:one_to_many}, the human recording is weighted 0.22 and each teacher 0.13, so the human recording receives about 1.7 times the weight of any single teacher. Appendix~\ref{app:teacher_weights} compares other weight allocations.

\section{Experimental Setup}
\label{sec: experiment}

\subsection{Training Settings}
\label{sec:training_details}

BELLE and MELLE share the same backbone and are each trained on three corpora: LS, LS+6TTS, and the ParallelVox training set. Both use every realization of a text as a training target with the weights of Eq.~\eqref{eq:one_to_many}, which reduces to standard training on LS, where \(N=1\). To compare real and synthetic parallel speech at equal volume, BELLE is also trained on \textbf{LS+6TTS-881h}, a subset of LS+6TTS that matches the ParallelVox training set in hours; the subset is sampled by text, so every retained text keeps all seven realizations, and it is trained with the ParallelVox budget. BELLE-stream is initialized from BELLE trained on LS+6TTS and further trained on the same corpus with \(S_{\mathrm{text}}=20\) and \(S_{\mathrm{audio}}=50\) (\(\approx 0.8\)~s per audio chunk). Model and training details are in Appendix~\ref{app: modelconfig} and \ref{app: trainingdetails}.

\subsection{Evaluation Settings}
\label{sec: evaluation}

Intelligibility, speaker similarity, and quality are evaluated on the LibriSpeech test-clean subset, restricted to utterances of 4 to 10~s, with an open-source zero-shot TTS evaluation protocol \citep{Lee_evaluate-zero-shot-tts_2024}. Two inference conditions are used. In \textbf{Continuation}, the first 3~s of an utterance and its transcription serve as the prompt, and the model generates the rest of the utterance. In \textbf{Cross-Sentence}, an utterance of a speaker and its transcription serve as the prompt, and the model generates a different sentence in the voice of that speaker.

Intelligibility is measured by WER-H, the WER of a HuBERT-based \citep{hsu2021hubert} automatic speech recognition (ASR) model on the generated speech. Speaker similarity (SIM-o) is the cosine similarity between the WavLM-TDCNN \citep{chen2022wavlm} speaker embeddings of the generated speech and of the original speech prompt. For subjective evaluation, a mean opinion score (MOS) for overall quality and a similarity MOS (SMOS) for speaker similarity to the prompt are collected on a crowd-sourcing platform (Appendix~\ref{MOS}).

Diversity is evaluated on the ParallelVox single-speaker repeat test set (Sec.~\ref{sec:multivox}). Under the Cross-Sentence condition, each system synthesizes every prompt three times, and the three outputs form one group. Each sample is represented by the mean-pooled and L2-normalized hidden states of layer 13 of WavLM-Large \citep{chen2022wavlm}, a layer chosen following the layer-wise probing analysis of \citet{chiu2025probe}. Diversity (Cos.) is the mean pairwise cosine distance within a group, averaged over all groups, and the Ground Truth value is computed in the same way over the repeated human readings of each text. On these readings, this distance correlates with how much duration, F0, and energy vary from one reading to another (Appendix~\ref{app:diversity_validation}). Generated samples with a WER above 5\% are dropped before this computation, since their distance reflects generation errors rather than acoustic variation. Appendix~\ref{app: evaluation} gives further details.

\section{Results}
\subsection{Main Results}
\label{sec:main_results}

\textbf{BELLE outperforms MELLE and is competitive with larger TTS systems.} BELLE and MELLE share the backbone, the training corpus, and the training schedule and differ in the Sampling Module and the loss that trains it (Sec.~\ref{sec:evidential_sampling}), so their gap measures the effect of the evidential formulation. BELLE outperforms MELLE on nearly every metric under all three training corpora (Table~\ref{tab:main}; Appendix~\ref{app:significance}), including LS, which has one recording per text, so the evidential formulation helps even without parallel speech. Against these TTS systems, which have up to 40 times more parameters or 50 times more training data (Table~\ref{tab:external}), BELLE gives the best Continuation results, ranks third on both Cross-Sentence metrics, and has the lowest RTF while supporting streaming generation.

\begin{table*}[t]
  \centering
  \caption{Controlled comparison under a shared training and evaluation pipeline. WER-H and SIM-o are measured on LibriSpeech test-clean, and Cos.\ ($\times 10^{-4}$) on the ParallelVox single-speaker repeat test set. MOS and SMOS were collected only for Ground Truth and the models trained on LS+6TTS.}
  \label{tab:main}
  \small
  \setlength{\tabcolsep}{4pt}
    \begin{tabular}{@{}llccccccc@{}}
      \toprule
      \multirow{2}{*}{\textbf{System}} & \multirow{2}{*}{\textbf{Training data}} & \multicolumn{2}{c}{\textbf{Continuation}} & \multicolumn{2}{c}{\textbf{Cross-Sentence}} & \multicolumn{1}{c}{\textbf{Diversity}} & \multicolumn{2}{c}{\textbf{Subjective}} \\ \cmidrule(r){3-4} \cmidrule(r){5-6} \cmidrule(r){7-7} \cmidrule(r){8-9}
      & & WER-H$\downarrow$ & SIM-o$\uparrow$ & WER-H$\downarrow$ & SIM-o$\uparrow$ & Cos.$\uparrow$ & MOS$\uparrow$ & SMOS$\uparrow$ \\ \midrule
      Ground Truth & -                      & 2.15 & 0.668 & 2.15 & 0.672 & 46.7 & 4.20 & 3.81 \\ \midrule
      VALL-E       & ParallelVox            & 7.46 & 0.340 & 16.3 & 0.452 & 17.4 & -    & - \\
      MELLE        & ParallelVox            & 3.06 & 0.406 & 5.60 & 0.538 & 19.9 & -    & - \\
      BELLE        & ParallelVox            & 3.00 & 0.449 & 5.73 & 0.570 & 21.5 & -    & - \\
      BELLE        & LS+6TTS-881h  & 2.57 & 0.510 & 4.05 & 0.634 & 20.3 & -    & - \\ \midrule
      MELLE        & LS            & 2.78 & 0.452 & 8.76 & 0.561 & 10.4 & -    & - \\
      BELLE        & LS            & 2.39 & 0.457 & 5.01 & 0.576 & 17.3 & -    & - \\ \midrule
      MELLE        & LS+6TTS       & 2.59 & 0.488 & 3.83 & 0.606 & 21.3 & 4.02 & 3.80 \\
      BELLE        & LS+6TTS       & \textbf{2.13} & \textbf{0.519} & \textbf{2.99} & \textbf{0.641} & \textbf{23.9} & \textbf{4.21} & \textbf{4.13} \\ \bottomrule
    \end{tabular}

  \vspace{8pt}
  \centering
  \caption{Comparison with larger TTS systems on LibriSpeech test-clean. Data is the training volume. RTF is measured with batch size 1. Streaming indicates whether a system supports streaming generation. Llasa-8B uses direct inference. $^\dagger$Estimated from the reported architecture.}
  \label{tab:external}
  \small
  \setlength{\tabcolsep}{2.8pt}
    \begin{tabular}{@{}lcccccccc@{}}
      \toprule
      \multirow{2}{*}{\textbf{System}} & \multirow{2}{*}{\textbf{Params}} & \multirow{2}{*}{\textbf{Data}} & \multicolumn{2}{c}{\textbf{Continuation}} & \multicolumn{2}{c}{\textbf{Cross-Sentence}} & \multirow{2}{*}{\textbf{RTF}$\downarrow$} & \multirow{2}{*}{\textbf{Streaming}} \\ \cmidrule(r){4-5} \cmidrule(r){6-7}
      & & & WER-H$\downarrow$ & SIM-o$\uparrow$ & WER-H$\downarrow$ & SIM-o$\uparrow$ & & \\ \midrule
      Ground Truth                        & -                & -      & 2.15 & 0.668 & 2.15 & 0.672 & -    & - \\ \midrule
      VALL-E \citep{wang2023neural}       & 0.37B$^\dagger$  & 60k~h  & 3.8  & -     & 5.9  & -     & -    & \xmark \\
      VALL-E 2 \citep{chen2024valle2}     & -                & 50k~h  & 2.32 & 0.504 & \textbf{2.44} & 0.643 & - & \xmark \\
      RALL-E \citep{xin2024ralle}         & 0.37B$^\dagger$  & 44k~h  & -    & -     & 2.8  & 0.49  & -    & \xmark \\
      CLAM-TTS \citep{kimclam}            & 1.2B$^\dagger$   & 55k~h  & 2.36 & 0.477 & 5.11 & 0.495 & 0.42 & \xmark \\
      Llasa-8B \citep{ye2025llasa}        & 8B               & 250k~h & 2.29 & 0.483 & -    & -     & 0.79 & \xmark \\
      CosyVoice 2 \citep{du2024cosyvoice} & 0.65B            & 167k~h & -    & -     & 3.37 & \textbf{0.713} & 0.33 & \cmark \\ \midrule
      BELLE                               & 0.2B             & 4.8k~h & \textbf{2.13} & \textbf{0.519} & 2.99 & 0.641 & \textbf{0.28} & \cmark \\ \bottomrule
    \end{tabular}
\end{table*}

\textbf{Parallel realizations improve most objective metrics.} To isolate the effect of parallel realizations, BELLE is trained on LS and on LS+6TTS-881h, two corpora of similar volume built on LS that differ in whether each text has one realization or seven. With seven realizations per text, four of the five objective metrics improve, namely Cross-Sentence WER-H, both SIM-o scores, and diversity, whereas Continuation WER-H becomes worse. Several realizations of a text thus supply the spread that the evidential loss estimates.

\textbf{Real and synthetic parallel speech are comparable, and synthetic speech scales further.} To compare the two sources at equal volume, BELLE trained on ParallelVox is compared with BELLE trained on LS+6TTS-881h under the same training budget. Each model leads on the test set closer to its own training domain: the synthetic corpus, built on LS, gives lower WER-H and higher SIM-o on LibriSpeech test-clean, whereas ParallelVox gives higher diversity on its own single-speaker repeat test set. Neither source dominates, so the two support BELLE at a comparable level. They differ in cost: ParallelVox reaches 881~h only after five public corpora are merged, whereas synthetic realizations can be generated for any corpus with one recording per text at the cost of TTS inference. Scaling LS+6TTS-881h to the full LS+6TTS improves every objective metric and yields the strongest BELLE in Table~\ref{tab:main}.

\subsection{Where the Gain over MELLE Comes From}
\label{sec:mechanism}

\textbf{BELLE learns an input-dependent variance.} The end-task metrics show that BELLE outperforms MELLE but do not show where the gain comes from. To test whether it comes from the per-frame variance that BELLE predicts from the input (Sec.~\ref{sec:evidential_sampling}) and which the KL loss of MELLE pulls toward one for every frame, the per-frame standard deviation \(\sigma_t\) is exported from BELLE and MELLE trained on LS+6TTS on 196 Cross-Sentence utterances. Because the two models place \(\sigma_t\) on different scales, only relative quantities are compared. MELLE's \(\sigma_t\) is almost constant, as Eq.~\eqref{eq:melle_kl} predicts, with a coefficient of variation (CV) of 6\% within an utterance and 1.2\% across utterances. BELLE's \(\sigma_t\) varies much more, with a CV of 19\% within an utterance and 7.0\% across utterances, so its predicted variance changes with the input both from frame to frame and from sentence to sentence. Both models assign a larger \(\sigma_t\) to frames with fast spectral change, such as onsets, offsets, and formant transitions, than to steady frames of similar energy (Appendix~\ref{app:variance_analysis}), but the increase is 17\% for BELLE, in every analyzed utterance, and only 5\% for MELLE.

\textbf{The learned variance follows real speech.} To test whether the variation in BELLE's \(\sigma_t\) reflects real speech, BELLE trained on ParallelVox is teacher-forced on one reading of each sentence in the single-speaker repeat test set, and its predicted \(\sigma_t^2\) is compared frame by frame with the variance across the repeated human readings of that sentence. Over 2{,}185 sentences, the two are positively correlated (median Spearman \(\rho = 0.18\)). Frames with fast spectral change, low energy, or an irregular alignment could raise both quantities at once, so the correlation is recomputed after these three factors are removed from both, and it stays at a median of 0.17, positive in 91\% of the sentences. When the predicted variance of one sentence is instead compared with the readings of a different sentence, the correlation drops to about zero (Appendix~\ref{app:variance_analysis}), so it reflects the content of each sentence rather than the analysis procedure. Spectral flux alone is not correlated with the variance across readings (median \(\rho = -0.06\)), so a predictor based on spectral change would not reproduce this result. The predicted variance of BELLE also responds more strongly than that of MELLE: the slope of \(\log \sigma_t^2\) on the logarithm of the variance across readings is 0.069 for BELLE and 0.024 for MELLE. Together with the previous analysis, this ties the gain over MELLE to the input-dependent variance of Sec.~\ref{sec:evidential_sampling}.

\subsection{Why the NIG Prior Fits Speech}
\label{sec:why_nig}

BELLE places an NIG prior on every mel frame (Eq.~\eqref{eq:nig_prior_mel_vector}) instead of a Gaussian with a fixed or a free variance. To test whether speech calls for this choice, the repeated readings of each text in the ParallelVox single-speaker repeat test set are aligned by dynamic time warping (DTW), which gives several observations of every frame without any model (Appendix~\ref{app:variance_analysis}). These observations show two properties, each of which matches one part of the NIG prior and rules out a simpler distribution.

\textbf{(i) Some frames vary several times more than others.} Within an utterance, the standard deviation across repeated readings is about 3 times larger at the most variable frames than at the most stable ones (99th vs.\ 1st percentile), whereas sampling noise alone would give 1.3 times. A fixed variance, as in MELLE, cannot represent this difference, whereas BELLE predicts the spread of each frame from the input.

\textbf{(ii) The variance of a frame is uncertain.} A text has only a few readings, so the variance of a frame computed from them carries large sampling noise: most of its variation from frame to frame is noise rather than a real difference in spread (Appendix~\ref{app:variance_analysis}). A free per-frame variance, as learned by a Gaussian NLL loss, is fit directly to these noisy values and therefore fits the noise as well. The Inverse-Gamma prior over \(\sigma_t^2\) in BELLE instead treats the variance as uncertain and pulls its estimate toward the prior mean, so the noise in a few readings moves the estimate less. Consistent with this, a Gaussian NLL loss with a free per-frame variance, trained in place of MELLE's KL loss, trails both MELLE and BELLE (Appendix~\ref{app:hetero}). Integrating over the uncertain variance gives the Student-\(t\) predictive distribution of Eq.~\eqref{eq:student_t_predictive_mel_vector}, whose tails are heavier than those of a Gaussian, and the real readings are themselves heavier-tailed than a Gaussian once the small number of readings and the alignment are accounted for (Appendix~\ref{app:variance_analysis}).

\textbf{Modeling these properties raises the held-out likelihood.} To test whether the two properties hold beyond the texts they are measured on, the densities are fit on half of the texts and evaluated on the other half. Each held-out reading is predicted from the other readings of its text. Relative to a Gaussian with one global variance, a Gaussian whose variance at each frame and band is estimated from the other readings and pulled toward the global variance, as in (i) and (ii), raises the held-out log-likelihood by 0.033 nats per observation, and replacing it with the Student-\(t\) predictive distribution of an uncertain variance raises it by another 0.066 nats (Appendix~\ref{app:variance_analysis}).

\subsection{Streaming Generation}
\label{sec:stream}

\begin{wraptable}{r}{0.4\textwidth}
  \centering
  \small
  \vspace{-12pt}
  \caption{Cross-Sentence results of streaming systems: SMLLE \citep{sun2025zero}, IST-LM \citep{yang2024interleaved}, and BELLE-stream trained on LS+6TTS.}
  \label{tab:stream}
  \vspace{-4pt}
  \begin{tabular}{@{}lcc@{}}
        \toprule
        \textbf{System} & WER-H$\downarrow$ & SIM-o$\uparrow$ \\ \midrule
        SMLLE        & 6.37          & 0.489          \\
        IST-LM       & 4.53          & \textbf{0.653} \\ \midrule
        BELLE-stream & \textbf{4.01} & 0.557          \\ \bottomrule
      \end{tabular}
  \vspace{-10pt}
\end{wraptable}

As a continuous-valued AR model that samples each frame in a single pass, BELLE can generate speech chunk by chunk while the text is still arriving (Appendix~\ref{app:streaming}). Among the streaming systems in Table~\ref{tab:stream}, BELLE-stream has the lowest WER-H, and its SIM-o lies between those of SMLLE and IST-LM. Because BELLE-stream conditions each audio chunk only on the text received so far (Eq.~\eqref{eq:stream_factorization}), its WER-H is higher than that of non-streaming BELLE. On one NVIDIA H200 GPU with batch size 1, BELLE-stream runs at a real-time factor (RTF) of 0.308 with a median first-packet latency of 232~ms, so it synthesizes speech faster than real time.

\section{Conclusion}
We propose BELLE, a TTS model that explicitly models the ``one-to-many'' distribution with evidential deep learning, predicting an NIG prior over the mean and variance of each mel-spectrogram frame from the input. We construct two parallel-speech corpora, ParallelVox from real human recordings and LS+6TTS from six pre-trained TTS teachers, which provide the multiple realizations per text that this variance estimation requires. Our experiments lead to three findings. First, BELLE outperforms MELLE under the same training conditions, is competitive with larger TTS systems on the objective metrics of LibriSpeech test-clean, and supports streaming generation. Second, real speech has the properties that the NIG prior assumes, and the variance predicted by BELLE changes with the input and follows the variation across repeated human readings of a text. Third, parallel speech improves BELLE, and at the same volume, synthetic parallel speech performs comparably to real parallel speech and is far easier to scale.

\subsection*{AI use statement}
We used large language models (LLMs) to polish the writing of our paper and to help edit code. The research ideas, method design, experiments, and analyses are our own. We reviewed all LLM-assisted text and code and take full responsibility for the content of the paper.

\subsection*{Ethics statement}
BELLE is a zero-shot TTS model: given a few seconds of a speaker's voice, it generates new speech in that voice. The same capability that serves assistive communication and spoken dialogue systems can be misused for impersonation, fraud, and the spread of fabricated speech. Detectors of synthetic speech, audio watermarking, and consent from the speakers whose voices are used as prompts reduce these risks, and we encourage their use together with models of this kind.

All human recordings used in this paper come from public corpora: LibriSpeech and the five sources of ParallelVox, whose licenses are listed in Table~\ref{tab:multivox_sources}. LS+6TTS adds speech synthesized by six publicly released TTS models in the voices of LibriSpeech speakers, and this synthetic speech is used only to train and evaluate the models in this paper. The subjective evaluation was run with crowd-sourced raters who were paid for their participation (Appendix~\ref{MOS}).

\subsection*{Reproducibility statement}
Code and audio samples are available at the links given in the abstract. The model architecture is described in Sec.~\ref{sec:architecture} and Appendix~\ref{app: modelconfig}, the Sampling Module and its losses in Sec.~\ref{sampling}, and the derivation of the evidential loss in Appendix~\ref{app:edl}. ParallelVox is built only from public corpora, and Appendix~\ref{app:multivox} gives its sources, preprocessing, filtering thresholds, and train/test split. LS+6TTS is built from LibriSpeech with six public TTS models, and Appendix~\ref{app: training data} gives the preprocessing, the teachers, and how they are prompted. Training hyperparameters, including the optimizer, batch size, number of updates, learning-rate schedule, and loss weights of every model, are listed in Appendix~\ref{app: experiment}. The evaluation follows an open-source protocol, and Appendix~\ref{app: evaluation} lists the ASR and speaker-verification models used, the subjective evaluation procedure, and the computation of the diversity metric. Appendix~\ref{app:significance} describes the significance tests, and Appendix~\ref{app:variance_analysis} describes the variance analyses of Sec.~\ref{sec:mechanism} and Sec.~\ref{sec:why_nig}.

\bibliography{iclr2027_paper}
\bibliographystyle{iclr2027_conference}

\newpage
\appendix

\section{Limitations}
\label{sec: limitation}

BELLE is trained and evaluated only on English speech, so its behavior on other languages and in cross-lingual synthesis is untested.

The official implementation of MELLE is not public, so every MELLE result in this paper comes from our re-implementation, trained with the same backbone, data, and training schedule as BELLE. Its absolute scores can therefore differ from those reported by \citet{meng2024autoregressive}.

The comparison of realization weights (Appendix~\ref{app:teacher_weights}) and the ablations of the Sampling Module and the flux loss (Appendices~\ref{app:ablation} and~\ref{app:hetero}) are run on corpora built from a 72~h subset of LS (Appendix~\ref{app: smalldetails}). Whether the same rankings hold at the scale of LS+6TTS is not tested.

The largest training corpus in this paper is LS+6TTS with 4.8k~h, whereas the larger TTS systems in Table~\ref{tab:external} are trained on 44k to 250k~h. How BELLE scales beyond this volume, for example when parallel speech is synthesized for a larger corpus with one recording per text, is not tested.

\section{Bayesian Interpretation of BELLE}
\label{app:bayesian_clarification}

Bayesian neural networks (BNNs) place priors over the network weights. BELLE follows evidential deep learning (EDL; \citealp{amini2020deep}) instead: the network weights are point estimates, and the prior is placed over the parameters of the likelihood of each frame.

\subsection{Empirical Bayes and Type-II Maximum Likelihood}
For one mel band of one frame, the observation \(y\) follows a Gaussian with mean \(\mu\) and variance \(\sigma^2\). BELLE treats \(\boldsymbol{\theta} = (\mu, \sigma^2)\) as a random variable and predicts the parameters \(\boldsymbol{m} = (\gamma, \nu, \alpha, \beta)\) of an NIG prior over \(\boldsymbol{\theta}\) from the input \(\boldsymbol{x}\). Training maximizes the marginal likelihood of the observation,
\begin{equation}
    \log p(y \mid \boldsymbol{m}) = \log \int p(y \mid \boldsymbol{\theta})\, p(\boldsymbol{\theta} \mid \boldsymbol{m}) \, d\boldsymbol{\theta},
\end{equation}
in which \(\boldsymbol{\theta}\) is integrated out in closed form (Appendix~\ref{app:edl}). Fitting the parameters of a prior by maximizing the marginal likelihood is known as empirical Bayes, or type-II maximum likelihood. In contrast, a regression loss fits a point estimate of \(y\), and the KL loss of MELLE fits one Gaussian per frame whose variance is pulled toward one (Eq.~\eqref{eq:melle_kl}).

\subsection{Sampling at Generation Time}
At generation time, each frame is sampled from the predicted NIG prior in three steps (Eq.~\eqref{eq: invgamma sample}): a variance \(\boldsymbol{\sigma}_t^2\) is drawn from the Inverse-Gamma distribution \(\Gamma^{-1}(\boldsymbol{\alpha}_t, \boldsymbol{\beta}_t)\), a mean \(\boldsymbol{\mu}_t\) from \(\mathcal{N}(\boldsymbol{\gamma}_t, \boldsymbol{\sigma}_t^2/\boldsymbol{\nu}_t)\), and the frame \(\boldsymbol{z}_t\) from \(\mathcal{N}(\boldsymbol{\mu}_t, \boldsymbol{\sigma}_t^2)\). The sample \(\boldsymbol{z}_t\) therefore follows the Student-\(t\) distribution of Eq.~\eqref{eq:student_t_predictive_mel_vector}, and its spread changes with the input through \((\boldsymbol{\nu}_t, \boldsymbol{\alpha}_t, \boldsymbol{\beta}_t)\). Replacing \(\boldsymbol{z}_t\) with the location \(\boldsymbol{\gamma}_t\) at inference raises WER-H (Appendix~\ref{app:hetero}), so the sampled spread affects the generated speech.

A model trained only with an L1 or L2 loss predicts the conditional median or mean of a frame, which averages over the valid realizations of a text. Sampling from the NIG prior instead draws one realization, with a larger spread at frames with fast spectral change (Sec.~\ref{sec:mechanism}). Consistent with this, BELLE reaches a higher diversity than MELLE under all three training corpora (Table~\ref{tab:main}).

\subsection{Comparison with Gaussian Sampling}
\label{app:edl_vs_gaussian}

The Sampling Modules of BELLE and MELLE differ in the family of the predictive distribution and in how its variance is trained.

\paragraph{Predictive distribution.}
Integrating out the NIG prior gives a Student-\(t\) predictive distribution with \(2\alpha\) degrees of freedom \citep{amini2020deep}. As \(\alpha \to \infty\), the Student-\(t\) approaches a Gaussian, so the predictive distribution of BELLE contains the Gaussian of MELLE as a limiting case and can also represent heavier tails. Real repeated readings have tails heavier than a Gaussian once the small number of readings and the alignment are accounted for (Appendix~\ref{app:variance_analysis}).

\paragraph{Variance.}
In the KL loss of MELLE (Eq.~\eqref{eq:melle_kl}), each variance term \(\sigma_{t,d}^2 - \log \sigma_{t,d}^2\) is minimized at \(\sigma_{t,d}^2 = 1\) for every frame, which pulls the model toward homoscedasticity, a variance that is the same for every input. The evidential loss has no term that pulls the variance toward a fixed value, so the variance is fit to the observed frames and can change with the input, which is known as heteroscedasticity. The predicted variances of the two models follow this difference: the coefficient of variation of \(\sigma_t\) within an utterance is 19\% for BELLE and 6\% for MELLE (Sec.~\ref{sec:mechanism}).

\section{Modeling Temporal and Spectral Dependencies}
\label{app:dependency_modeling}

The NIG prior of Eq.~\eqref{eq:nig_prior_mel_vector} is placed on each frame and acts on each mel band separately, whereas neighboring mel frames are strongly correlated, and so are neighboring bands within a frame. Correlations across frames are modeled exactly by the AR factorization, and the only independence assumption is that the bands of one frame are independent given its hidden representation.

\paragraph{Across frames.}
BELLE does not treat frames as independent. The factorization of Eq.~\eqref{eq:ar_factorization} is the chain rule of probability and holds for any joint distribution of the frames, so the NIG prior of frame \(t\) models the conditional distribution \(p(\boldsymbol{y}_t \mid \boldsymbol{y}_{<t}, \boldsymbol{x})\) rather than a marginal distribution of the frame. The AR LM computes \(\boldsymbol{e}_t\) from the text and all previous frames (Sec.~\ref{sec:architecture}), and the NIG parameters of frame \(t\) are a function of \(\boldsymbol{e}_t\). A change in any previous frame therefore changes both the location \(\boldsymbol{\gamma}_t\) and the spread of frame \(t\). At inference, each generated frame is fed back as input to the next step, so the dependence of every frame on the frames before it is kept during generation.

\paragraph{Across bands.}
Within frame \(t\), the \(D\) bands are modeled as independent given \(\boldsymbol{e}_t\), and correlations across bands are still captured in two ways. First, all bands share \(\boldsymbol{e}_t\), and their NIG parameters are predicted together by one linear layer, so the locations \(\boldsymbol{\gamma}_t\) of all bands, which carry the spectral envelope of the frame, are predicted jointly. The independence applies only to the random deviation of each band around its location. Second, this deviation passes through two networks that act on all bands together before the output is formed: the denoising MLP maps all \(D\) bands of \(\boldsymbol{z}_t\) to the frame \(\boldsymbol{y}_t^{(1)}\), and the Postnet convolves over neighboring frames with all bands as input channels. Both networks are trained with the regression loss of Eq.~\eqref{eq:regression_loss_belle_rhs} against the ground-truth frames, so they can remove deviations that are inconsistent across bands.

\paragraph{Why the bands are not modeled jointly.}
A joint prior over the \(D\) bands would require a full covariance matrix per frame, with \(D(D+1)/2 = 3{,}240\) variance parameters for \(D = 80\) instead of 80, and sampling from its conjugate prior, the Normal-Inverse-Wishart distribution, requires drawing an \(80 \times 80\) covariance matrix and factorizing it at every frame. The factorization over bands keeps the Sampling Module to one linear layer and the sampling of Eq.~\eqref{eq: invgamma sample} to element-wise operations. MELLE makes the same assumption with its diagonal Gaussian (Eq.~\eqref{eq:melle_param}), so the comparison between BELLE and MELLE differs in the prior over each band, not in how the bands are coupled.

\section{Derivation of the Evidential Loss}
\label{app:edl}

We drop the indices \(t\) and \(d\) and consider one mel band of one frame. The observation \(y\) and its NIG prior are
\begin{equation}
  y \sim \mathcal{N}(\mu, \sigma^2),\quad \mu \sim \mathcal{N}\left(\gamma, \frac{\sigma^2}{\nu}\right),\quad \sigma^2 \sim \Gamma^{-1}(\alpha, \beta),
\end{equation}
where \(\Gamma^{-1}(\alpha, \beta)\) is the Inverse-Gamma distribution, and the parameters \(\boldsymbol{m} = (\gamma, \nu, \alpha, \beta)\) satisfy \(\gamma \in \mathbb{R}\), \(\nu > 0\), \(\alpha > 1\), and \(\beta > 0\). The joint density of \((\mu, \sigma^2)\) is the NIG density
\begin{equation}
  p(\mu, \sigma^2 \mid \boldsymbol{m}) = \frac{\beta^\alpha \sqrt{\nu}}{\Gamma(\alpha)\sqrt{2\pi\sigma^2}} \left(\frac{1}{\sigma^2}\right)^{\alpha+1} \exp\left\{ -\frac{2\beta + \nu(\gamma - \mu)^2}{2\sigma^2} \right\},
\end{equation}
where \(\Gamma(\cdot)\) is the gamma function. The marginal likelihood of \(y\) integrates out both parameters:
\begin{equation}
  p(y \mid \boldsymbol{m}) = \int_{0}^{\infty} \int_{-\infty}^{\infty} p(y \mid \mu, \sigma^2)\, p(\mu, \sigma^2 \mid \boldsymbol{m})\, \mathrm{d}\mu\, \mathrm{d}\sigma^2.
\end{equation}
Integrating out \(\mu\) first gives \(y \mid \sigma^2 \sim \mathcal{N}(\gamma, \sigma^2(1+\nu)/\nu)\). Integrating out \(\sigma^2\) under the Inverse-Gamma prior then gives a Student-\(t\) distribution,
\begin{equation}
  p(y \mid \boldsymbol{m}) = \mathrm{St}\left(y;\, \gamma,\, \frac{\beta(1+\nu)}{\nu\alpha},\, 2\alpha\right),
\end{equation}
with location \(\gamma\), squared scale \(\beta(1+\nu)/(\nu\alpha)\), and \(2\alpha\) degrees of freedom. The negative logarithm of this density is \(\mathcal{L}_{\text{NLL}}\) of Eq.~\eqref{eq:edl_loss_mel_vector}, with \(\Omega = 2\beta(1+\nu)\).

The regularizer \(\mathcal{L}_{\text{R}} = |y - \gamma|\,(2\nu + \alpha)\) follows \citet{amini2020deep}. The quantity \(2\nu + \alpha\) is the evidence, the number of virtual observations that the prior represents, and multiplying it by the error lowers the confidence of the prior on frames that are predicted poorly. The two terms are combined with the weight \(\lambda\) in Eq.~\eqref{eq:edl_loss_mel_vector}.

\section{Model and Training Details}
\label{app: experiment}

\subsection{Model Configuration}
\label{app: modelconfig}

The Prenet is a 3-layer network with a dropout rate of 0.5, which is kept active at inference as in Tacotron \citep{wang2017tacotron}. The AR LM is a decoder-only Transformer with 12 blocks, 16 attention heads, a hidden size of 1024, a feed-forward dimension of 4096, and a dropout rate of 0.1\footnote{Our code is modified from \url{https://github.com/lifeiteng/vall-e}.}. The Sampling Module is one linear layer that outputs the NIG parameters, and the denoising MLP has three residual layers. The Postnet consists of five convolutional blocks with a kernel size of 5 and 256 channels. Waveforms are synthesized with a pretrained HiFi-GAN vocoder \citep{kong2020hifigan}\footnote{\url{https://huggingface.co/mechanicalsea/speecht5-tts}}.

\subsection{Training of the Main Models}
\label{app: trainingdetails}

BELLE and MELLE are trained on LS, LS+6TTS, and the ParallelVox training set with the weighted loss of Eq.~\eqref{eq:one_to_many}. On LS+6TTS, the human recording of a text is weighted 0.22 and each teacher realization 0.13 in Eq.~\eqref{eq:one_to_many} (Appendix~\ref{app:teacher_weights}); on ParallelVox, all recordings of a text are weighted equally; on LS, each text has one recording. Training uses AdamW with a batch of about 160K frames on 16 NVIDIA A800 GPUs for 450K updates. The learning rate rises linearly to \(5\times10^{-4}\) over the first 10\% of the updates and then decays linearly to zero. The models trained on ParallelVox and on LS+6TTS-881h share a budget of about 440K updates (Sec.~\ref{sec:training_details}). For BELLE, \(\lambda = 0.5\) in Eq.~\eqref{eq:edl_loss_mel_vector}, and \(\lambda_{\text{samp}} = 0.2\) and \(\lambda_{\text{flux}} = 0.5\) in Eq.~\eqref{eq:total_loss_belle}. For MELLE, \(\lambda_{\text{samp}} = 0.1\) and \(\lambda_{\text{flux}} = 0.5\), as in the original paper \citep{meng2024autoregressive}.

BELLE-stream is initialized from BELLE \textit{LS+6TTS} and trained on LS+6TTS with a batch of 80K frames on 8 NVIDIA A800 GPUs for 150K updates, with \(\lambda_{\text{flux}} = 0.1\) and all other hyperparameters as in BELLE. A text chunk contains \(S_{\mathrm{text}} = 20\) phonemes after G2P conversion, and an audio chunk contains \(S_{\mathrm{audio}} = 50\) mel frames, about 0.8~s. Utterances whose ratio of mel frames to phonemes is below 2.5 are removed from training, so that each text chunk has enough audio frames to be generated from.

VALL-E \citep{wang2023neural} is trained from scratch on the ParallelVox training set as a discrete-token baseline. The recordings of a text are treated as independent text--recording pairs, so ParallelVox is used as a standard TTS corpus.

\subsection{Training of the Analysis Models}
\label{app: smalldetails}

The weight comparison of Appendix~\ref{app:teacher_weights} and the ablations of Appendices~\ref{app:ablation} and~\ref{app:hetero} use smaller corpora built from the LibriSpeech \texttt{train-clean-100} subset. Processing this subset as LS (Appendix~\ref{app: training data}) gives 72~h of speech, denoted \textit{LS-100}. \textit{LS-100+2TTS} adds the realizations of XTTS-v2 and MaskGCT, for 218~h in total, and \textit{LS-100+6TTS} adds those of all six teachers, for 455~h in total. Both are trained with the weighted loss of Eq.~\eqref{eq:one_to_many}, a batch of about 80K frames, and 47K updates.

\section{Parallel Realizations under the AR Factorization}
\label{app:ar_realizations}

BELLE models the distribution of each frame given the text and the previous frames (Eq.~\eqref{eq:ar_factorization}), and during training each realization is teacher-forced on its own previous frames. The \(N\) realizations of a text \(\boldsymbol{x}\) therefore give one observation of each of \(N\) different conditional distributions \(p(\boldsymbol{y}_t \mid \boldsymbol{y}^{(i)}_{<t}, \boldsymbol{x})\), not \(N\) observations of one conditional distribution. They are nonetheless \(N\) observations of the same distribution \(p(\boldsymbol{y} \mid \boldsymbol{x})\), and training with Eq.~\eqref{eq:one_to_many} maximizes a weighted likelihood of these observations.

\paragraph{Inputs at training and inference.}
During training, the input of BELLE is the text of an utterance and the target is its mel-spectrogram; no prompt is given. The prompts used by the teachers to build LS+6TTS (Appendix~\ref{app: training data}) only determine which voice each teacher synthesizes and are never seen by BELLE, so every realization of a text, human or synthetic, is a sample of \(p(\boldsymbol{y} \mid \boldsymbol{x})\), which includes variation in speaker and prosody. At inference, the prompt text is concatenated with the input text, the prompt speech is given as the first frames, and the model continues generation from it. The prompt speech therefore acts as the prefix \(\boldsymbol{y}_{<t}\) and sets the voice of the generated speech.

\paragraph{The loss decomposes over realizations.}
Under the model, the realizations of a text are independent given \(\boldsymbol{x}\), because every frame of every realization draws its own mean and variance from the NIG prior (Eq.~\eqref{eq: invgamma sample}). The log-likelihood of the \(N\) realizations is therefore
\begin{equation}
  \log p_\theta\big(\boldsymbol{y}^{(1)}, \ldots, \boldsymbol{y}^{(N)} \mid \boldsymbol{x}\big)
  = \sum\nolimits_{i=1}^{N} \log p_\theta\big(\boldsymbol{y}^{(i)} \mid \boldsymbol{x}\big)
  = \sum\nolimits_{i=1}^{N} \sum\nolimits_{t=1}^{T_i} \log p_\theta\big(\boldsymbol{y}^{(i)}_t \mid \boldsymbol{y}^{(i)}_{<t}, \boldsymbol{x}\big),
\end{equation}
where \(T_i\) is the number of frames of realization \(i\) and each term is the Student-\(t\) likelihood of Eq.~\eqref{eq:student_t_predictive_mel_vector}. Computing \(\mathcal{L}_{\text{NLL}}\) on all realizations of a text jointly and summing the losses computed on each realization separately thus give the same objective, and the regularizer \(\mathcal{L}_{\text{R}}\) is also a sum over realizations. Eq.~\eqref{eq:one_to_many} is this sum with the weights \(w_i\), and with equal weights it is the maximum-likelihood objective for \(N\) observations of \(p(\boldsymbol{y} \mid \boldsymbol{x})\). If all realizations of a text instead shared one draw of the variance, the joint likelihood would not factorize in this way.

\paragraph{What the realizations estimate.}
Let \(p_{\text{data}}(\boldsymbol{y} \mid \boldsymbol{x})\) be the distribution from which the realizations of \(\boldsymbol{x}\) are drawn. As the number of realizations grows, the average of this objective approaches the expected log-likelihood under \(p_{\text{data}}\), and maximizing it minimizes
\begin{equation}
  D_{\text{KL}}\big(p_{\text{data}}(\boldsymbol{y} \mid \boldsymbol{x}) \,\|\, p_\theta(\boldsymbol{y} \mid \boldsymbol{x})\big)
  = \sum\nolimits_{t} \mathbb{E}_{\boldsymbol{y}_{<t} \sim p_{\text{data}}}\Big[ D_{\text{KL}}\big(p_{\text{data}}(\boldsymbol{y}_t \mid \boldsymbol{y}_{<t}, \boldsymbol{x}) \,\|\, p_\theta(\boldsymbol{y}_t \mid \boldsymbol{y}_{<t}, \boldsymbol{x})\big)\Big]
\end{equation}
by the chain rule of the KL divergence. The conditional distribution of each frame, including its spread, is therefore fit on average over the prefixes that realizations of \(\boldsymbol{x}\) produce. One recording provides one such prefix per text, whereas \(N\) realizations provide \(N\) prefixes that share the linguistic content and differ in voice and prosody. The AR LM computes \(\boldsymbol{e}_t\) with the same parameters for every prefix, so realizations with similar prefixes inform the same region of the conditional distribution, in the same way that an AR language model learns next-token distributions although most of its prefixes occur only once in the training data. At the first frame, all realizations share the context \(\boldsymbol{x}\) and are \(N\) observations of the same conditional distribution \(p(\boldsymbol{y}_1 \mid \boldsymbol{x})\).

\paragraph{Empirical check.}
The second analysis of Sec.~\ref{sec:mechanism} teacher-forces BELLE \textit{ParallelVox} on one reading of each sentence and compares its predicted variance with the variance across the repeated readings at the same frames. The two are positively correlated (median Spearman \(\rho = 0.18\)), also after spectral flux, energy, and alignment irregularity are controlled for (median partial \(\rho = 0.17\)), and the correlation vanishes when the predicted variance of one sentence is paired with the readings of another (Appendix~\ref{app:variance_analysis}). The conditional spread that BELLE learns from realizations with different prefixes thus follows the spread across real readings of a sentence.

\section{Parallel Speech Corpora}
\label{app:corpora}

\subsection{ParallelVox}
\label{app:multivox}

ParallelVox merges five public English corpora in which the same text is recorded more than once, by different speakers or by one speaker on different occasions (Table~\ref{tab:multivox_sources}). All audio is downmixed to mono, resampled to 16~kHz, and trimmed with Silero VAD \citep{SileroVAD}, and each text is converted to phonemes with the espeak-ng G2P backend, which uses 78 phoneme symbols. Each text is stored once with all of its recordings, and in training all recordings of a text enter Eq.~\eqref{eq:one_to_many} with equal weights. A text has 2 to 91 recordings in the training set and 2 to 18 in the test set.

\begin{table}[h]
  \centering
  \caption{Source corpora of ParallelVox after VAD trimming and before WER filtering.}
  \label{tab:multivox_sources}
  \small
  \setlength{\tabcolsep}{4pt}
  \begin{tabular}{@{}lrrrl@{}}
    \toprule
    \textbf{Source} & \textbf{Recordings} & \textbf{Hours} & \textbf{Speakers} & \textbf{License} \\
    \midrule
    LibriHeavy \citep{kang2024libriheavy} & 312,959 & 1,017.7 & 3,392 & CC BY 4.0 \\
    Parallel Audiobook Corpus & 84,791 & 134.4 & 53 & Public Domain \\
    VCTK 0.92 \citep{yamagishi2019vctk} & 34,981 & 20.7 & 109 & CC BY 4.0 \\
    \begin{tabular}[t]{@{}l@{}}CMU ARCTIC \citep{kominek2004cmu} +\\ L2-Arctic \citep{zhao2018l2arctic}\end{tabular} & 18,625 & 15.5 & 42 & Mixed \\
    \begin{tabular}[t]{@{}l@{}}Common Voice 17.0 \citep{ardila2020common},\\ parallel subset\end{tabular} & 284 & 0.2 & 255 & CC-0 \\
    \midrule
    \textbf{Total} & \textbf{451,640} & \textbf{1,188.5} & \textbf{3,851} & \\
    \bottomrule
  \end{tabular}
\end{table}

\paragraph{Quality filtering.}
Each recording is transcribed with Whisper-large-v2 \citep{radford2023robust} using greedy decoding and scored with UTMOS22-Strong \citep{saeki2022utmos}. A recording is dropped when its WER or MOS fails the threshold, and a text is kept only if at least two of its recordings remain, since parallel speech needs at least two realizations of a text. ParallelVox uses only the filter WER~\(\leq\)~10\%, which keeps 123,640 texts, 344,303 recordings, and 903~h. Table~\ref{tab:multivox_filter} shows how stricter WER or MOS thresholds reduce the corpus.

\begin{table}[h]
  \centering
  \caption{Size of ParallelVox under different WER and MOS thresholds. The first row is the setting used in this paper.}
  \label{tab:multivox_filter}
  \small
  \begin{tabular}{@{}ccrrr@{}}
    \toprule
    \textbf{WER $\leq$} & \textbf{MOS $\geq$} & \textbf{Texts} & \textbf{Recordings} & \textbf{Hours} \\
    \midrule
    10\% & --- & 123,640 & 344,303 & 903.0 \\
    10\% & 3.8 & 60,012 & 175,356 & 400.1 \\
    10\% & 4.0 & 38,618 & 114,741 & 244.4 \\
    5\% & --- & 92,734 & 262,337 & 638.5 \\
    3\% & --- & 73,683 & 214,831 & 474.9 \\
    \bottomrule
  \end{tabular}
\end{table}

\paragraph{Train and test split.}
\label{app:multivox_split}
Texts whose recordings all come from one speaker form the single-speaker repeat test set: 3,306 texts, 6,831 recordings, 22~h, and 250 speakers. The remaining 120,334 texts, each recorded by at least two speakers, form the training set: 337,472 recordings, 881~h, and 3,406 speakers. The test set records how one speaker varies across repeated readings of the same text, which makes it the human reference for the diversity metric (Appendix~\ref{app:diversity_eval}).

\subsection{LS+6TTS}
\label{app: training data}

LS is built from the LibriSpeech training set \citep{panayotov2015librispeech}. Long silences are removed with Silero VAD \citep{SileroVAD}, and utterances of 0.5 to 14~s are kept, which gives 706~h. Audio is resampled to 16~kHz and converted to 80-band mel-spectrograms, and transcripts are converted to phonemes with espeak-ng\footnote{\url{https://github.com/espeak-ng/espeak-ng}}.

Six pretrained TTS teachers synthesize every text of LS: CosyVoice2 \citep{du2024cosyvoice}, an LLM with a flow-matching decoder; IndexTTS \citep{deng2025indextts}, an AR model over discrete acoustic tokens; SparkTTS \citep{wang2025spark}, a single-stage LLM-based model; F5TTS \citep{chen2024f5}, a flow-matching model; MaskGCT \citep{wang2024maskgct}, a masked generative non-autoregressive model; and XTTS-v2 \citep{casanova2024xtts}, a multilingual model. Each teacher is prompted with a different utterance of the same speaker, drawn at random from the LS recordings of that speaker with a duration of 3 to 8~s, and its output is resampled to 16~kHz. With the human recording, each text has seven realizations, and the corpus totals 4,817~h.

\subsection{Weights of Human and Synthetic Realizations}
\label{app:teacher_weights}

In Eq.~\eqref{eq:one_to_many}, the human recording of an LS+6TTS text is weighted 0.22 and each of the six teacher realizations 0.13, so the weights sum to one and the human recording counts about 1.7 times as much as any single teacher. Table~\ref{tab:weight_ablation} compares this choice with three other allocations that also sum to one, on \textit{LS-100+6TTS} (Appendix~\ref{app: smalldetails}). The weights 0.22/0.13 give the lowest WER-H in both conditions, whereas 0.16/0.14 gives the highest SIM-o in both. Raising the human weight above 0.22 worsens all four metrics. The SIM-o differences between 0.16/0.14 and 0.22/0.13 are at most 0.006, whereas the WER-H differences are 0.07 and 0.44, so 0.22/0.13 is used.

\begin{table}[h]
  \centering
  \caption{Weights of the human recording and of each teacher realization in Eq.~\eqref{eq:one_to_many}, on \textit{LS-100+6TTS}. The second row is the setting used in this paper.}
  \label{tab:weight_ablation}
  \small
    \begin{tabular}{cccccc}
      \toprule
      \multirow{2}{*}{\textbf{Human}} & \multirow{2}{*}{\textbf{Each teacher}} & \multicolumn{2}{c}{\textbf{Continuation}} & \multicolumn{2}{c}{\textbf{Cross-Sentence}} \\ \cmidrule(r){3-4} \cmidrule(r){5-6}
       &  & WER-H $\downarrow$ & SIM-o $\uparrow$ & WER-H $\downarrow$ & SIM-o $\uparrow$ \\ \midrule
      0.16 & 0.14 & 2.64 & \textbf{0.415} & 5.17 & \textbf{0.453} \\
      0.22 & 0.13 & \textbf{2.57} & 0.409 & \textbf{4.73} & 0.449 \\
      0.28 & 0.12 & 2.78 & 0.406 & 5.47 & 0.444 \\
      0.34 & 0.11 & 2.76 & 0.405 & 5.47 & 0.446 \\ \bottomrule
    \end{tabular}
\end{table}

\section{Evaluation Details}
\label{app: evaluation}

\subsection{Objective Evaluation}

All models are evaluated on the same set of LibriSpeech test-clean utterances of 4 to 10~s, following the protocol of VALL-E \citep{wang2023neural} as implemented by \citet{Lee_evaluate-zero-shot-tts_2024}. Intelligibility is measured with two ASR models, a Conformer-Transducer\footnote{\url{https://huggingface.co/nvidia/stt_en_conformer_transducer_xlarge}} \citep{gulati20conformer} and a fine-tuned HuBERT-Large\footnote{\url{https://huggingface.co/facebook/hubert-large-ls960-ft}} \citep{hsu2021hubert}, whose WERs are denoted WER-C and WER-H. Speaker similarity is the cosine similarity between WavLM-TDCNN\footnote{\url{https://github.com/microsoft/UniSpeech/tree/main/downstreams/speaker_verification}} \citep{chen2022wavlm} speaker embeddings of the generated speech and of the prompt: SIM-o uses the original prompt, and SIM-r uses the prompt reconstructed by the vocoder. The main text reports WER-H and SIM-o; WER-C and SIM-r appear in Table~\ref{tab:abl1}. BELLE-stream is evaluated only under the Cross-Sentence condition.

\subsection{Subjective Evaluation}
\label{MOS}

\begin{figure}[h]
  \centering
  \begin{subfigure}[b]{0.48\linewidth}
    \centering
    \includegraphics[width=\linewidth]{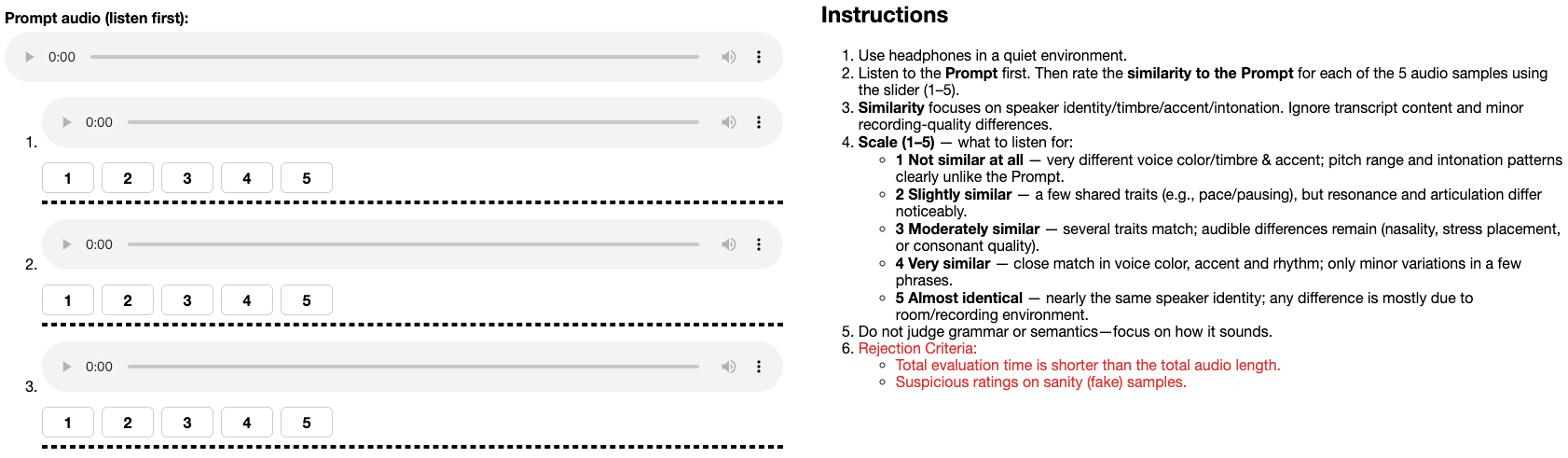}
    \caption{Speaker similarity (SMOS)}
    \label{fig:smos}
  \end{subfigure}
  \hfill
  \begin{subfigure}[b]{0.48\linewidth}
    \centering
    \includegraphics[width=\linewidth]{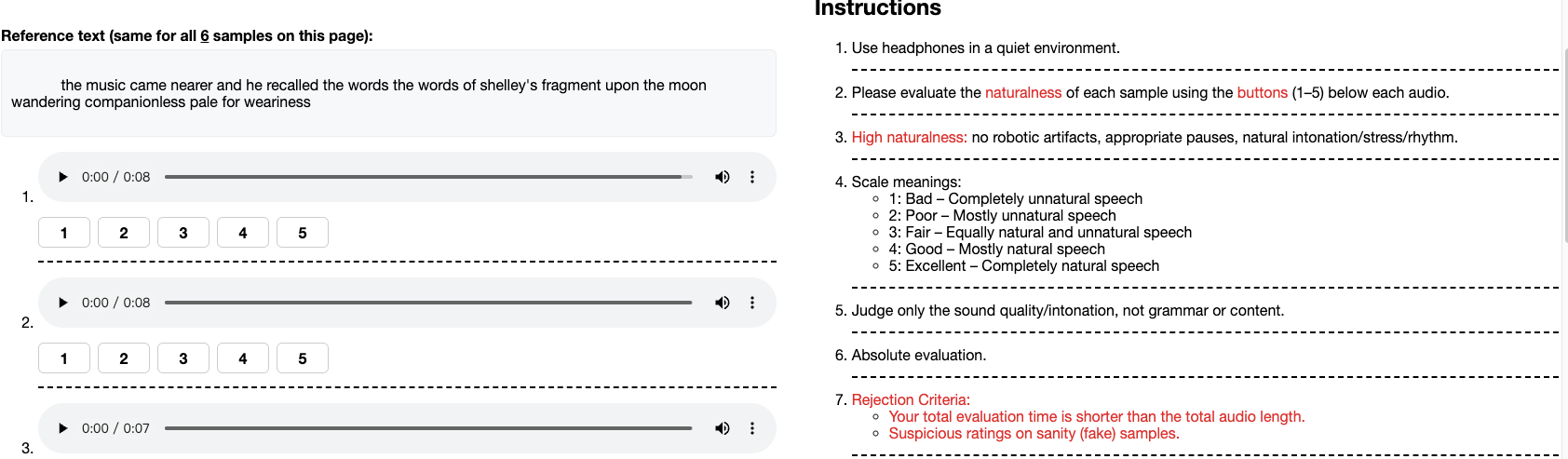}
    \caption{Speech quality (MOS)}
    \label{fig:mos}
  \end{subfigure}
  \caption{Instructions shown to the raters in the subjective evaluation.}
  \label{fig:subjective_evaluation}
\end{figure}

MOS and SMOS are collected for Ground Truth, MELLE \textit{LS+6TTS}, and BELLE \textit{LS+6TTS}. One Cross-Sentence sample is selected per speaker, which gives 40 samples per system, and all samples are resampled to 16~kHz. For MOS, raters score quality and naturalness on a 1-to-5 scale and are told to ignore differences in style, such as timbre, emotion, and prosody. For SMOS, each sample is paired with the reference utterance, and raters score how similar the speaker identity (timbre) is on a 1-to-5 scale while ignoring differences in content, grammar, and audio quality. The order of the samples is shuffled for each rater. Fifteen native speakers rated the samples on Amazon Mechanical Turk, and every rater scored both BELLE and MELLE. Figure~\ref{fig:subjective_evaluation} shows the instructions. The participants received about \$100 in total. Part of the samples can be heard at \url{https://windowso.github.io/BELLE-demo/}.

\subsection{Diversity Metric}
\label{app:diversity_eval}

Diversity is measured on the single-speaker repeat test set of ParallelVox (Appendix~\ref{app:multivox_split}). Of its 3,306 texts, 3,270 have a Cross-Sentence prompt, a different utterance of the same speaker of 3 to 8~s, and are used. Each system synthesizes each text three times from the same prompt, so the three outputs of a group differ only through sampling. Each waveform is passed through WavLM-Large, and the hidden states of layer 13 (of 24) are mean-pooled over time, excluding padding, and L2-normalized; Appendix~\ref{app:diversity_validation} explains this choice. The distance between two outputs \(a\) and \(b\) is \(1 - |\cos(a, b)|\), where the absolute value keeps two embeddings that point in opposite directions from counting as maximally different. The diversity of a group is the mean distance over all pairs in it, and Cos.\ is the mean over groups, reported \(\times 10^{-4}\) in Table~\ref{tab:main}. The Ground Truth value applies the same computation to the human readings of each text.

Generated samples with a WER above 5\% are dropped before the distances are computed, since the distance of an unintelligible sample reflects generation errors rather than acoustic variation, and a group with fewer than two remaining samples has no pair to compare and is left out of the mean.

\subsection{Validity of the Diversity Metric}
\label{app:diversity_validation}

The choice of layer follows the probing analysis of \citet{chiu2025probe}, which finds that early layers of self-supervised speech models encode basic acoustics, that middle layers combine them into more abstract traits, and that intermediate representations track prosody better than dedicated speaker embeddings. Layer 13 of WavLM-Large is therefore used instead of a speaker-embedding model.

To test whether the metric reflects how differently a sentence is spoken, five prosodic quantities are measured on every sample: duration, median F0, F0 standard deviation, mean energy, and energy standard deviation. For each group, the variation of each quantity across the samples is correlated with the WavLM distance of the group. Table~\ref{tab:diversity_validation} reports Spearman's \(\rho\) over the groups of each group source, and a permutation test gives \(p \le 0.004\) in every cell.

\begin{table}[h]
  \centering
  \caption{Spearman's \(\rho\) between the within-group variation of prosodic quantities and the within-group WavLM distance.}
  \label{tab:diversity_validation}
  \small
  \setlength{\tabcolsep}{4pt}
  \begin{tabular}{@{}lccccc@{}}
    \toprule
    \textbf{Group source} & \textbf{Duration} & \textbf{Median F0} & \textbf{F0 s.d.} & \textbf{Energy} & \textbf{Energy s.d.} \\ \midrule
    Ground Truth (human readings)     & \textbf{0.54} & \textbf{0.70} & \textbf{0.61} & 0.21 & \textbf{0.65} \\
    BELLE \textit{ParallelVox}        & 0.07 & 0.37 & 0.25 & \textbf{0.34} & 0.33 \\
    BELLE \textit{LS+6TTS}            & 0.19 & 0.38 & 0.25 & 0.29 & 0.29 \\ \bottomrule
  \end{tabular}
\end{table}

All 15 correlations are positive, and on four of the five quantities the correlation is strongest on the human readings, where the speaker and the text are fixed and only the reading varies. The metric therefore tracks how prosody varies across repeated readings of one sentence, which is the variation that diversity is meant to measure.

\section{Additional Results}
\label{app:additional_experiments}

\subsection{Statistical Significance}
\label{app:significance}

BELLE and MELLE are scored on the same test items, so each objective difference is tested with a paired bootstrap over items, namely sentences for WER-H and SIM-o and prompt groups for Cos., and a paired Wilcoxon test on the same items gives the same conclusions. MOS and SMOS are tested with a paired Wilcoxon test over the rated samples, since every rater scored both systems. On LS, all five objective differences favor BELLE with \(p \le 0.004\), and on LS+6TTS, all seven differences favor BELLE with \(p < 0.05\). On ParallelVox, the two SIM-o differences and the diversity difference favor BELLE with \(p \le 0.007\), whereas the two WER-H differences are not significant (Table~\ref{tab:significance}).

\begin{table}[h]
  \centering
  \caption{BELLE \(-\) MELLE differences and \(p\)-values, from a paired bootstrap for the objective metrics and a paired Wilcoxon test for MOS and SMOS. MOS and SMOS were collected only on LS+6TTS.}
  \label{tab:significance}
  \small
  \setlength{\tabcolsep}{4pt}
  \begin{tabular}{@{}lccc@{}}
    \toprule
    \textbf{Metric} & \textbf{ParallelVox} & \textbf{LS} & \textbf{LS+6TTS} \\ \midrule
    Continuation WER-H    & \(-0.07\) (\(p=0.52\))   & \(-0.39\) (\(p<0.001\))  & \(-0.46\) (\(p<0.001\)) \\
    Continuation SIM-o    & \(+0.044\) (\(p<0.001\)) & \(+0.005\) (\(p=0.004\)) & \(+0.031\) (\(p<0.001\)) \\
    Cross-Sentence WER-H  & \(+0.13\) (\(p=0.56\))   & \(-3.75\) (\(p<0.001\))  & \(-0.84\) (\(p<0.001\)) \\
    Cross-Sentence SIM-o  & \(+0.032\) (\(p<0.001\)) & \(+0.015\) (\(p<0.001\)) & \(+0.036\) (\(p<0.001\)) \\
    Diversity (Cos.)      & \(+1.67\) (\(p=0.007\))  & \(+6.90\) (\(p<0.001\))  & \(+2.60\) (\(p<0.001\)) \\
    MOS                   & --                       & --                       & \(+0.19\) (\(p=0.021\)) \\
    SMOS                  & --                       & --                       & \(+0.33\) (\(p<0.001\)) \\ \bottomrule
  \end{tabular}
\end{table}

\subsection{Details of the Variance Analyses}
\label{app:variance_analysis}

\paragraph{Predicted variance.}
For the first analysis of Sec.~\ref{sec:mechanism}, the per-frame \(\sigma_t\) of BELLE \textit{LS+6TTS} and MELLE \textit{LS+6TTS} is exported during inference on 196 Cross-Sentence utterances of LibriSpeech test-clean. For BELLE, the variance is the analytic variance of the Student-\(t\) predictive distribution, \(\beta(1+\nu)/((\alpha-1)\nu)\), computed from the predicted NIG parameters rather than from samples; for MELLE, it is \(\exp(\log\sigma^2)\) from the predicted log-variance. The per-frame \(\sigma_t\) is the square root of this variance averaged over the 80 mel bands. The CV within an utterance is the standard deviation of \(\sigma_t\) over the frames of the utterance divided by their mean, averaged over utterances, and the CV across utterances is computed over the per-utterance means of \(\sigma_t\).

The spectral change of a frame is its spectral flux, the absolute difference between consecutive log-mel frames averaged over the 80 bands. Within each utterance, the 20\% of frames with the highest flux form the fastest group, the 20\% with the lowest flux form the slowest group, and the ratio of their mean \(\sigma_t\) is computed. Frames with fast spectral change, such as onsets and offsets, are often quieter than steady frames, and quiet frames have noisier log-mel values, so a larger \(\sigma_t\) at these frames could reflect low energy rather than spectral change. The comparison is therefore made among frames of similar energy: the energy of a frame is the mean of its 80 log-mel values, the frames of each utterance are split into five equal-frequency energy bins, the ratio is computed within each bin and averaged over bins, and a bin is skipped when either group has fewer than three frames. The per-utterance ratios are averaged over utterances, with bootstrap confidence intervals. With the ground-truth recordings as the AR input (teacher forcing), the increase is 12\% for BELLE and 4\% for MELLE.

\paragraph{Alignment of readings.}
The readings of a text are aligned on their mel-spectrograms by DTW. The per-band mean of each reading is first subtracted, which removes offsets in timbre and level that would otherwise distort the alignment, since the DTW cost is the mean absolute mel difference. The medoid of the group, the reading with the smallest summed DTW distance to the others, provides the reference time axis, and every other reading is aligned to it under a Sakoe--Chiba band of 25\% of the length and at least 10 frames. The variance across readings is then computed at every frame of the medoid.

\paragraph{Correlation with the variance across readings.}
BELLE \textit{ParallelVox} and MELLE \textit{ParallelVox} are teacher-forced on the medoid of each group of the single-speaker repeat test set, and their predicted \(\sigma_t^2\) is compared with the variance across readings at the same frames. The analysis uses the 2{,}185 groups whose readings are all at most 14~s long, the longest utterance length used in training. For each group, the Spearman correlation is computed over frames, and its significance is assessed against the correlations obtained under all circular shifts of the predicted variance, which keep the autocorrelation of both sequences, with a false discovery rate correction over groups. As a control, the predicted variance of one sentence is paired with the readings of another sentence. The partial correlation removes from both variables the ranks of spectral flux, energy, and alignment irregularity, defined as the fraction of alignment paths that enter a medoid frame through a non-diagonal step, averaged over readings and smoothed over 5 frames. The local DTW cost is not used as a control, because it is computed from the same frame differences as the variance across readings. The amplitude of the response is the slope of \(\log \sigma_t^2\) on the logarithm of the variance across readings within each group, which does not depend on the overall scale of \(\sigma_t^2\).

For BELLE, the median correlation is 0.18 (95\% CI [0.17, 0.18]); it is positive in 88\% of the groups and significant in 57\%. The median partial correlation is 0.17 (95\% CI [0.16, 0.17]); it is positive in 91\% of the groups and significant in 59\%. Within bins of spectral flux and energy, the median correlation is 0.16. On the 92 groups with at least three readings, the median correlation is 0.19 and the median partial correlation 0.20. In the mismatched control, the median correlation is 0.01 and no group is significant, and spectral flux itself has a median correlation of \(-0.06\) with the variance across readings. The median slope is 0.069 for BELLE and 0.024 for MELLE, with a paired difference of 0.044 (95\% CI [0.042, 0.046]) that is positive in 89\% of the groups.

\paragraph{Properties of real speech tested in Sec.~\ref{sec:why_nig}.}
These statistics use the aligned human readings alone, without any model. For each text, the standard deviation of a frame is the square root of the variance across readings averaged over the 80 bands, as for \(\sigma_t\) in Sec.~\ref{sec:mechanism}, and the ratio of its 99th to its 1st percentile over the frames of the text is computed. The median ratio is 3.2 (interquartile range [2.7, 3.6]) over the 114 texts with at least three readings and 2.5 over the 10 texts with at least six readings. If the variance were the same at every frame, sampling noise alone would give median ratios of 1.30 and 1.15.

The reliability of the per-frame variance is measured with a shrinkage estimator, which writes the variance of a frame as a weighted average of the global variance and the empirical variance of the frame across readings, with the weight chosen to minimize the mean squared error. With three readings of a text, the best weight puts about 79\% on the global variance and 21\% on the empirical variance of the frame. Because this weight equals the share of sampling noise in the variation of the empirical variance across frames, about 79\% of this variation is sampling noise, and this share decreases as the number of readings grows. The Inverse-Gamma prior gives an estimate of the same form: for \(n\) observations with a known mean and the prior \(\sigma^2 \sim \Gamma^{-1}(\alpha,\beta)\), the posterior mean of \(\sigma^2\) is \(w\,\beta/(\alpha-1) + (1-w)\,s^2\), where \(s^2\) is the empirical variance and \(w = (\alpha-1)/(\alpha-1+n/2)\), so \(\alpha\) sets how far the estimate moves toward the prior mean.

For the tails, each observation is compared with the other readings of its frame. For a text with \(n\) readings, its deviation from the mean of the other readings, divided by their standard deviation and by \(\sqrt{1+1/(n-1)}\), follows a Student-\(t\) distribution with \(n-2\) degrees of freedom when the readings are Gaussian. Passing this value through the cumulative distribution function of that Student-\(t\) and then through the inverse standard normal gives a value that is standard normal under a Gaussian, whatever \(n\) is. On the 114 texts with at least three readings, the probabilities that this value exceeds 2.58 and 3.29 in absolute value are 1.43 (95\% CI [1.26, 1.59]) and 1.50 (95\% CI [1.33, 1.67]) times their nominal values of 1\% and 0.1\%. To account for the alignment, Gaussian readings are generated around the medoid, warped along the real alignment paths, and realigned by DTW; whether their per-frame variance is the empirical variance of the data, a smoothed version of it, or a shrunk version of it, the two ratios stay between 1.01 and 1.04. The excess on real readings remains on the half of the frames with the lowest spectral flux, with ratios of 1.55 and 1.65, and on the texts with at least six readings, with ratios of 1.22 and 1.61.

For the held-out log-likelihood, the 114 texts of the single-speaker repeat test set with at least three readings are split by text into two halves of 57. Each observation, one mel band of one frame of one reading, is predicted from the other readings of its text, and all three densities use the mean of these readings as their location. The first Gaussian uses one global variance for all observations. The second uses the variance of the other readings at the same frame and band, pulled toward the global variance with a weight of 0.30 on the local variance. The Student-\(t\) uses a scale derived from a shrunk local variance and a fitted degree of freedom. The global variance, the weights, the scale factor, and the degree of freedom are fit on one half, and the log-likelihood is computed on the other half. The second Gaussian gains 0.033 nats per observation over the first, and the Student-\(t\) gains another 0.066 nats, which sums to 7.9 nats per frame over its 80 bands.

\subsection{Ablation of the Sampling Module and the Flux Loss}
\label{app:ablation}

Table~\ref{tab:abl1} removes either the Sampling Module or the flux loss from BELLE \textit{LS-100+2TTS} (Appendix~\ref{app: smalldetails}). Removing the Sampling Module raises WER-H from 2.66 to 5.00 in Continuation and from 4.84 to 13.07 in Cross-Sentence, whereas removing the flux loss raises it to 3.25 and 6.37. All eight metrics follow the same order, with the full model best and the model without the Sampling Module worst, so both components help and the Sampling Module accounts for the larger share of the gain.

\begin{table*}[h]
  \centering
  \caption{Ablation of the Sampling Module and the flux loss on BELLE \textit{LS-100+2TTS}. \cmark\ indicates enabled, and \xmark\ indicates disabled.}
  \label{tab:abl1}
  \small
    \begin{tabular}{@{}cccccccccc@{}}
      \toprule
      \multirow{2}{*}{\textbf{Sampling}} &
      \multirow{2}{*}{\textbf{Flux loss}} &
      \multicolumn{4}{c}{\textbf{Continuation}} &
      \multicolumn{4}{c}{\textbf{Cross-Sentence}} \\ \cmidrule(r){3-6} \cmidrule(r){7-10}
      &        & WER-C & WER-H & SIM-r & SIM-o & WER-C & WER-H & SIM-r & SIM-o \\ \midrule
      \xmark & \cmark & 4.30  & 5.00  & 0.382 & 0.344 & 12.04 & 13.07 & 0.392 & 0.351 \\
      \cmark & \xmark & 2.50  & 3.25  & 0.388 & 0.347 & 5.61  & 6.37  & 0.417 & 0.372 \\
      \cmark & \cmark & \textbf{2.04} & \textbf{2.66} & \textbf{0.398} & \textbf{0.362} & \textbf{4.12} & \textbf{4.84} & \textbf{0.433} & \textbf{0.395} \\ \bottomrule
    \end{tabular}
\end{table*}

\subsection{Free-Variance Gaussian Baseline and Inference-Time Sampling}
\label{app:hetero}

The Sampling Module of MELLE already predicts a per-frame variance (Eq.~\eqref{eq:melle_param}), and only the KL loss of Eq.~\eqref{eq:melle_kl} pulls this variance toward one. To test whether a free variance alone accounts for the gain of BELLE, the KL loss is replaced by the Gaussian NLL
\begin{equation}
  \mathcal{L}_{\text{GNLL}} = \sum\nolimits_{t=1}^{T}\sum\nolimits_{d=1}^{D}\left(\frac{(y_{t,d}-\mu_{t,d})^2}{2\sigma_{t,d}^2} + \frac{1}{2}\log\sigma_{t,d}^2\right),
\end{equation}
which has no term toward a fixed value and no prior over the variance. The four systems in Table~\ref{tab:hetero} are trained on \textit{LS-100+2TTS} and differ only in the Sampling Module and its loss; the first and last rows are taken from Table~\ref{tab:abl1}.

\begin{table}[h]
  \centering
  \caption{Sampling Modules and their losses on \textit{LS-100+2TTS}.}
  \label{tab:hetero}
  \small
  \begin{tabular}{@{}lcc@{}}
    \toprule
    \textbf{Sampling Module and loss} & \textbf{Continuation WER-H}$\downarrow$ & \textbf{Cross-Sentence WER-H}$\downarrow$ \\ \midrule
    None & 5.00 & 13.07 \\
    Gaussian, KL toward unit variance (MELLE) & 3.24 & 6.83 \\
    Gaussian, NLL with free variance & 3.88 & 7.31 \\
    NIG, evidential loss (BELLE) & \textbf{2.66} & \textbf{4.84} \\ \bottomrule
  \end{tabular}
\end{table}

The free-variance model trails MELLE in both conditions and trails BELLE by 2.5 points of Cross-Sentence WER-H, so a free variance alone does not account for the gain. Two properties of \(\mathcal{L}_{\text{GNLL}}\) are consistent with this result. Its gradient with respect to \(\mu_{t,d}\) is scaled by \(1/\sigma_{t,d}^2\), so frames that receive a large variance contribute little to learning the mean. It also fits the variance of each frame to the few realizations of a text: \textit{LS-100+2TTS} has three realizations per text, and on real speech with three readings per text, most of the frame-to-frame variation of such a variance is sampling noise (Appendix~\ref{app:variance_analysis}). The Inverse-Gamma prior of BELLE instead pulls this estimate toward the prior mean (Sec.~\ref{sec:why_nig}).

The sampled spread is also used at generation time. For BELLE \textit{LS-100+2TTS}, replacing the sample \(\boldsymbol{z}_t\) with the location \(\boldsymbol{\gamma}_t\) at inference raises Cross-Sentence WER-H from 4.84 to 6.16.

\section{Streaming Generation}
\label{app:streaming}

\paragraph{Why streaming.}
In a spoken dialogue system, the text to be spoken is often produced incrementally, for example by an LLM that generates one token at a time. A non-streaming TTS model starts synthesis only after the whole sentence is available, so the delay before the first audio grows with the length of the sentence. A streaming model starts from the first text chunk, so the first audio is available after a fixed delay.

\paragraph{Why AR models suit streaming.}
An AR model generates speech in time order, and each frame is conditioned only on the previous frames and the text (Eq.~\eqref{eq:ar_factorization}). A generated frame therefore never depends on later frames, and streaming only requires conditioning each audio chunk on the text received so far (Eq.~\eqref{eq:stream_factorization}). Models that generate all frames of an utterance jointly, such as diffusion and flow-matching TTS, must be modified to operate on chunks before they can stream; CosyVoice~2, for example, uses a chunk-aware flow-matching decoder \citep{du2024cosyvoice}.

\paragraph{Why BELLE suits streaming.}
BELLE generates mel-spectrogram frames directly, so each frame can be passed to the vocoder without an intermediate stage that decodes discrete tokens into acoustic features. Its Sampling Module predicts the NIG parameters of a frame with one linear layer and samples the frame in closed form (Eq.~\eqref{eq: invgamma sample}), so each frame costs one forward pass of the AR LM, the Sampling Module, and the denoising MLP. In contrast, continuous-valued AR models with diffusion or flow-matching heads \citep{liu2024autoregressive, wang2025felle, jia2025ditar} run several denoising steps for every frame. BELLE-stream therefore uses the same Sampling Module and the same evidential loss as BELLE, and only the input sequence changes.

\paragraph{Implementation.}
BELLE-stream interleaves text chunks and audio chunks as the input of the AR LM under a causal mask (Sec.~\ref{sec:architecture}), following the interleaving scheme of IST-LM \citep{yang2024interleaved}. Its chunk sizes and training settings are given in Appendix~\ref{app: trainingdetails}, and its results are in Sec.~\ref{sec:stream}.

\end{document}